\documentclass[usenatbib,onecolumn]{mn2e}
\usepackage{graphicx}
\usepackage{times}
\newcommand{\beq}{\begin{equation}}
\newcommand{\eeq}{\end{equation}}
\newcommand{\bea}{\begin{eqnarray}}
\newcommand{\eea}{\end{eqnarray}}
\def\lsim{\raise0.3ex\hbox{$\;<$\kern-0.75em\raise-1.1ex\hbox{$\sim\;$}}}
\def\gsim{\raise0.3ex\hbox{$\;>$\kern-0.75em\raise-1.1ex\hbox{$\sim\;$}}}
\title[Prompt Emission of High Energy Photons from Gamma Ray Bursts]
{Prompt Emission of High Energy Photons from Gamma Ray Bursts } 
\author[Nayantara Gupta and Bing Zhang]
{Nayantara Gupta\thanks{nayan@physics.unlv.edu} and 
Bing Zhang\thanks{bzhang@physics.unlv.edu}\\ 
Department of Physics and Astronomy, 
University of Nevada Las Vegas, Las Vegas, NV 89154, USA}
\begin{document}
\date{Accepted 2007; Received 2007; in original form 2007}
\pagerange{\pageref{firstpage}--\pageref{lastpage}} \pubyear{2007}
\maketitle
\label{firstpage}
\begin{abstract}
Within the internal shock scenario we consider different mechanisms of
high energy ($>1$ MeV) photon production inside a Gamma Ray Burst (GRB) 
fireball and derive the expected high energy photon spectra from individual
GRBs during the prompt phase. The photon spectra of leptonic and
hadronic origins are compared within different sets of parameter
regimes. Our results suggest that the high energy emission 
is dominated by the leptonic component if fraction of shock energy carried 
by electrons is not very small (e.g. $\epsilon_e > 10^{-3}$). For very small 
values of $\epsilon_e$ the hadronic emission component could be comparable
to or even exceed the leptonic component in the GeV-TeV regime. 
However, in this case a much larger energy budget of the fireball is 
required to account for the same level 
of the observed sub-MeV spectrum. The fireballs are therefore extremely
inefficient in radiation. For a canonical fireball bulk Lorentz factor  
(e.g. $\Gamma=400$), emissions above $\sim 10$ GeV are attenuated
by two-photon pair production processes. For a fireball with an even higher 
Lorentz factor, the cutoff energy is higher, and emissions of 10 TeV - PeV 
due to $\pi^0$-decay can also escape from the internal shocks. The flux
level is however too low to be detected by current TeV detectors, and 
these photons also suffer attenuation by external soft photons.  
GLAST LAT can detect prompt emission of bright long 
GRBs above 100 MeV. For short GRBs, the prompt emission can be only barely 
detected for nearby bright ones with relatively ``long'' durations
(e.g. $\sim$ 1 s). With the observed high energy spectrum alone, it 
appears that there is no clean picture to test the leptonic vs. hadronic
origin of the gamma-rays. Such an issue may be however addressed by
collecting both prompt and afterglow data. A moderate-to-high radiative
efficiency would suggest a leptonic origin of high energy photons, while
a GRB with an extremely low radiative efficiency but an extended high 
energy emission component would be consistent with (but not a proof for)
the hadronic origin.
\end{abstract}


\section{Introduction}
The study of Gamma Ray Bursts (GRBs) has been one of the most
interesting areas in astrophysics in the past few years. Ongoing
observational and theoretical investigations are disclosing the
physical origin, characteristics of these objects as well as bringing
new puzzles to us. EGRET detected high energy photons from five GRBs
coincident with triggers from the BATSE instrument \citep{jones}. GRB
940217 was detected by EGRET independent of BATSE trigger, which has
extended emission and with the highest energy photon of 18GeV \citep{hur}. 
\cite{gonzalez} discovered a distinct high energy component up
to 200 MeV in GRB 941017 that has a different temporal evolution with 
respect to the low energy component. Although even higher
energy gamma rays/neutrinos have not been firmly detected from GRBs 
yet, \cite{atkins} have provided tentative evidence of TeV 
emission from GRB 970417A. For a long time, GRBs have been identified 
as potential sources of ultrahigh energy cosmic rays \citep{wax1,vietri}. 
Within the standard fireball picture (e.g. M\'esz\'aros 2006), there are 
about a dozen mechanisms that can produce GeV-TeV gamma-rays from GRBs
(e.g. Zhang 2007). More theoretical and observational efforts are
needed to fully understand high energy emission
from GRBs.  From the theoretical aspect, it is essential to investigate 
the relative importance of various emission components to identify the
dominant mechanisms under certain conditions. 

The high energy photon spectra expected from GRBs during
the prompt and the afterglow phases have been derived by various
groups. In the scenario of external shock model the high energy photon
spectra during the early afterglow phase due to synchrotron and synchrotron
self Compton (SSC) emission by shock accelerated relativistic electrons and
protons have been studied 
\citep{mesz1,mesz2,pan1,lu1,tot1,der1,der2,der3,pan2,sari2,zhang1,fan3,gou}. 
In the case of a strong reverse shock emission component, the SSC emission 
in the reverse shock region or the crossing inverse Compton processes between 
the forward
and reverse shock regions are also important \citep{wang1,wang2,peer5}. The 
discovery of X-ray flares in early afterglows in the Swift era \citep{burrows}
also opens the possibility that scattering of the flaring photons from the
external shocks can give strong GeV emission \citep{wang3,fan2}.
The effect of cosmic infrared background on high energy delayed 
$\gamma$-rays from GRBs has been also widely discussed in the
literature \citep{dai,stecker,wang4,raz1,casa,murase}.
The most important high energy emission component is believed to be emitted 
from the prompt phase. Swift early X-ray afterglow data suggest that the GRB 
prompt emission is of ``internal'' origin, unlike the external-origin afterglow 
emission (Zhang et al. 2006, cf. Dermer 2007). 
The most widely discussed internal model of prompt
emission is the internal shock model \citep{rees}.
Within the internal shock model the spectrum of high
energy photons expected during the prompt phase has been studied
\citep{pilla,frag,bhat,raz1,peer1,peer2}. The various processes of
high energy photon production in the internal shocks are electron
synchrotron emission, SSC of electrons, synchrotron emission of protons, 
photon production through
$\pi^0$ decay produced in proton photon ($p\gamma$) interactions and
radiations by secondary positrons produced from $\pi^+$ decays. 
In this paper we consider all these processes self-consistently 
with a semi-analytical approach and study the relative importance of each
component within the internal shock scenario. The derived photon spectra 
are corrected for internal optical depth for pair production,
which is energy-dependent and also 
depends on various other parameters of GRBs e.g. their variability times, 
luminosities, the low energy photon spectra inside GRBs, and photon spectral 
break energies.
If the electrons cool down by synchrotron and SSC emission to trans-relativistic 
energies, then they accumulate near a value of Lorentz factor of around unity. 
The accumulated electrons affect the high energy photon spectrum by direct-Compton 
scattering and other processes, which make the spectrum significantly different 
from the broken power laws considered in this work, see \citep{peer4,peer2} for 
detailed discussions. In any case, for the values of parameters considered in 
the present paper this effect is not significant. 

GLAST's \citep{glast} burst monitor (GBM) will detect
photons in the energy range of 10keV to 25MeV and large area telescope
(LAT) will detect photons in the energy range of 20MeV and
1000GeV. With a large field of view  ($>2$ sr for LAT), GLAST will
detect high energy photons from many GRBs and open a new era
of studying GRBs in the high energy regime. This is supplemented by AGILE
\citep{agile}, which is designed to observe photons in the energy range of
10-40 keV and 30MeV-50GeV and also has a large field of view. There are several 
other ground based detectors
e.g. Whipple/VERITAS \citep{whipple}, Milagro \citep{milagro}, 
which have been searching or will search for $\sim$ TeV photons from GRBs. 
Detections or non-detections of high energy gamma rays from GRBs with space-based 
and ground-based detectors in the near future would make major steps in revealing 
the physical environment, bulk motion, mechanisms of particle acceleration and 
high energy photon production, photon densities, etc., of GRBs.

\section[]{Electron Synchrotron Radiation}
We define three reference frames: 
(i) the comoving frame or the wind rest frame is the rest frame of the outflowing 
ejecta expanding with a Lorentz factor $\Gamma$ with respect to the observer and
the central engine;
(ii) the source rest frame is attached to the GRB central engine at a redshift $z$;
and (iii) the observer's frame is the reference frame of the observer on earth, 
which is related to the source rest frame by the redshift correction factor.
We denote the quantities measured in the comoving frame with primes. 
The shock accelerated relativistic electrons lose energy by synchrotron radiation
and SSC in the shock region. Assuming a power law distribution of fresh electrons
accelerated from the internal shocks and considering a continuous injection of
electrons during the propagation of the shocks, the relativistic primary electron 
number distribution in the comoving frame can be expressed as a broken power law 
in energy \citep{sari1}
\beq
\frac{dN_e(E'_e)}{dE'_e}\propto\left\{ \begin{array}{l@{\quad \quad}l}
{{E'_e}^{-p}} & E'_{e,m}<E'_e<E'_{e,c}\\
{E'_e}^{-p-1} & E'_{e,c}<E'_e
\end{array}\right.
\eeq
in the case of slow cooling, where $E'_{e,m}$ is the minimum injection energy of 
electrons and $E'_{e,c}$ is the energy of an electron that loses its energy
significantly during the dynamic time scale, known as the cooling energy of 
the electrons. If the electrons are cooling fast
so that even the electrons with the minimum injection energy have cooled during
the dynamical time scale, by considering continuous injection of electrons from
the shock the comoving electron number distribution can be expressed as
\beq
\frac{dN_e(E'_e)}{dE'_e}\propto\left\{ \begin{array}{l@{\quad \quad}l}
{{E'_e}^{-2}} & E'_{e,c}<E'_e<E'_{e,m}\\
{E'_e}^{-p-1} & E'_{e,m}<E'_e
\end{array}\right.
\eeq
If the electrons cool down to sub-relativistic energies then they accumulate 
near electron Lorentz factor $\gamma'_e\sim 1$. This effect may distort the high 
energy photon spectrum by direct-Compton scattering \citep{peer4,peer2}, and
we focus on the parameter regime where this effect is not significant. 
The energies in the source rest frame and the comoving frame are related as $E_e
\simeq \Gamma E_e^{\prime}$, where $\Gamma$ is the average bulk Lorentz factor of 
the GRB fireball in the prompt phase. The expression for the minimum injection 
energy of electrons 
in the comoving frame is $E'_{e,m}=m_ec^2\bar{\gamma'_p}g(p)\frac{m_p}{m_e}
\frac{\epsilon_e}{\epsilon_p}$, where $g(p)=\frac{p-2}{p-1}$ for $p>>2$ and 
$g(p) \sim 1/6$ for $p=2$ \citep{RB}, $m_p$, $m_e$ are the masses of proton and 
electron, respectively, and $\bar{{\gamma'_p}}m_p c^2$ is the average internal
energy of protons in the comoving frame. We have assumed $\bar{{\gamma'_p}}$ to 
be of the order of unity (in principle $\bar{{\gamma'_p}}$ could be smaller than
unity). The total internal energy is distributed among electrons, 
protons and the internal magnetic fields within the internal shocks. The 
fractions of the total energy carried by electrons, protons and internal 
magnetic fields are represented by $\epsilon_e$, $\epsilon_p$ and $\epsilon_B$, 
respectively, where $\epsilon_e+\epsilon_p+\epsilon_B=1$. 
We have assumed that all the electrons and protons are accelerated in internal 
shocks. In reality, the shock accelerated particles may be only a fraction of
the total population and additional fractional parameters ($\xi_e$, $\xi_p$) 
may be introduced (e.g. Bykov \& M\'esz\'aros 1996). In such a case, the following
treatments are still generally valid by re-defining $\epsilon'_e=\epsilon_e/\xi_e$
and $\epsilon'_p=\epsilon_p/\xi_p$, while the relation $\epsilon_e+\epsilon_p+\epsilon_B
=1$ still holds.

The relativistic electrons lose their energy by synchrotron radiation and 
inverse Compton scattering \citep{pan1,sari2,zhang1}. 
The comoving cooling break energy in the relativistic electron spectrum can be 
derived by comparing the cooling and the dynamical time scales.
The comoving cooling time scale $t_{cool}^{\prime}$ of electrons is a convolution
of the cooling time scales for synchrotron radiation $t_{syn}^{\prime}$ 
and for inverse Compton (IC) scattering $t_{IC}^{\prime}$
\beq
\frac{1}{t_{cool}^{\prime}}=\frac{1}{t_{syn}^{\prime}}+\frac{1}{t_{IC}^{\prime}}~.
\eeq
We denote $U$ as the internal energy density of the internal shock, and $U_e$, 
$U_B$ as the energy densities of electrons and magnetic fields, respectively. 
The energy density of the synchrotron radiation is $U_{e,syn}=\frac{\eta_e U_e}
{1+Y_e}=\frac{\eta_e \epsilon_e U}{1+Y_e}$ \citep{sari2}, where the radiation
efficiency of electrons is $\eta_e=[(E_{e,c}^{\prime}/E_{e,m}^{\prime})^{2-p}, 
1]$ for slow and fast cooling, respectively, and 
\beq
Y_e=\frac{L_{e,IC}}{L_{e,syn}}=\frac{U_{e,syn}}{U_{B}}=
\frac{-1+\sqrt{1+4\eta_e\epsilon_e/\epsilon_B}}{2}
\eeq 
denotes the relative importance between the IC and the synchrotron emission
components\footnote{Strictly speaking, such a treatment is valid for the IC
process in the Thomson regime. However, this is also a reasonable approximation
if the peak of the spectral energy distribution of the IC component is in the
Thomson regime, which is generally the case for the calculations performed
in this paper.}. $L_{e,IC}$ and $L_{e,syn}$ are the luminosities of radiations 
 emitted in SSC and synchrotron emission of relativistic electrons respectively. 
The inverse of the cooling time scale of electrons can be expressed by the 
power divided by energy ($E_e^{\prime}=m_e\gamma_{e}^{\prime}c^2$),
\beq
\frac{1}{t_{cool}^{\prime}}=\frac{4}{3}\sigma_{e,T}{\beta_{e}^{\prime}}^2
\gamma_{e}^{\prime}\frac{c}{m_ec^2}(U_B+U_{e,syn}) =\frac{4}{3}\sigma_{e,T}
{\beta_{e}^{\prime}}^{2}\gamma_{e}^{\prime}\frac{cU\epsilon_B}{m_ec^2}
({1+Y_e})~,
\eeq
where $\sigma_{e,T}$ is Thomson cross-section of electrons, ${\beta_{e}}^{\prime} 
\simeq 1$ is the dimensionless speed of the relativistic electrons. The comoving
dynamical time scale is $t'_{dyn} \simeq \Gamma t_{v}$, where $\Gamma$ is the average 
Lorentz factor of the GRB, and $t_{v}$ is the variability time in the source rest 
frame of the GRB, which denotes the variability time scale of the central engine.
Throughout the paper, we assume that electron synchrotron radiation from the internal
shocks is the mechanism that power the prompt gamma-ray emission in the sub-MeV band.
However, for standard parameters 
within this scenario the cooling time scale of electrons is much shorter than the 
dynamical time scale of GRBs. As a result the flux density $\Big(E_{\gamma,s}
\frac{dN_{\gamma,s}(E_{\gamma,s})}{E_{\gamma,s}}\Big)$ below the cooling break 
energy is proportional to $E_{\gamma,s}^{-1/2}$ and cannot explain the 
harder spectral indices observed in many GRBs \citep{ghis}. If the magnetic field 
created by internal shocks decays on a length scale much shorter then the comoving 
width of the plasma, then the resulting synchrotron radiation can explain some of
the broadband GRB spectra observed by Swift \citep{peer3}. In this case 
the effective dynamical time scale is shorter by a factor of $f_c$ than its actual 
value. Hence, the ratio of the cooling and the dynamical time scale can be expressed as
\beq
\frac{t'_{dyn}}{t_{cool}^{\prime}}=f_c
\eeq
at the cooling energy $E_e^{\prime}=E_{e,c}^{\prime}$.
The expression of the electron cooling energy in the comoving frame can be written as
\beq
E_{e,c}^{\prime}= \gamma_{e,c}^{\prime}m_e c^2=m_ec^2\frac{3m_ec^2f_c}{4\Gamma t_{v}
\sigma_{e,T}cU\epsilon_B({1+Y_e})}=530{\rm keV}
\frac{t_{v,-2}\Gamma_{2}^5f_{c,2}}{L_{iso,51}\epsilon_{B,-1}({1+Y_e})}~.
\label{e_cool}
\eeq
Here and throughout the text the convention $Q_x=Q/10^x$ is adopted in cgs units.
In the above expression $L_{iso}$ is the luminosity corresponding to the 
energy $E_{iso}$ carried by all particles and the magnetic fields in the shocks. 
It is a fraction of the wind (outflow) luminosity $L_{iso}\sim \eta L_w$, where 
$\eta$ is the efficiency of converting the kinetic energy of the wind to 
the shock internal energy.
The luminosity $L_{iso}$ and internal energy $U$ are related as $U=L_{iso}/(4\pi\Gamma^2 
{r_{is}}^2 c)$, where $r_{is}=\Gamma^2 c t_v$ is the internal shock radius. 
The synchrotron spectrum is a multi-segment broken power law \citep{sari1} separated
by several breaks, including the emission frequency from electrons with the minimum 
injection energy, the cooling break frequency, and the synchrotron self-absorption
frequency \citep{ryb}. In the internal shocks, the magnetic field in the comoving frame 
can be expressed as \citep{bing3}
\beq
B^{\prime}\simeq 4.4\times10^{5} {\rm G} 
(\xi_{1}\epsilon_{B,-1})^{1/2} L_{iso,51}^{1/2} r_{is,13}^{-1} \Gamma_{2}^{-1}
=1.5\times10^{6} {\rm G} \frac{(\xi_{1} \epsilon_{B,-1} L_{iso,51})^{1/2}}
{{\Gamma}_{2}^{3} t_{v,-2}}
\label{B'}
\eeq
where $\xi$ is the compression ratio, which is about 7 for strong shocks.
The synchrotron self absorption energy ($E_{ssa}$) in internal shocks can be expressed 
as (Li \& Song 2004; Fan et al. 2005; cf. Pe'er \& Waxman 2004)
\beq
 E_{ssa}\simeq 0.24~{\rm keV} 
L_{\gamma,s,51}^{2/7}\Gamma_{2}^{3/7}r_{is,13}^{-4/7}{B_{5}^{\prime}}^{1/7}=0.69~{\rm keV} 
L_{iso,51}^{5/14} t_{v,-2}^{-5/7} {\Gamma_{2}}^{-8/7}(\xi_{1}\epsilon_{B,-1})^{1/14}
\Big(\frac{\epsilon_e\eta_e}{1+Y_e}\Big)^{2/7}
\eeq
where $L_{\gamma,s}=L_{iso}\epsilon_e\eta_e/(1+Y_e)$ is the isotropic gamma-ray luminosity
due to synchrotron radiation. 
The cooling break energy $E_{e,c}^{\prime}$ and 
the minimum injection energy $E_{e,m}^{\prime}$ of the electrons define two
break energies in the synchrotron photon spectrum.
The cooling break energy in the photon spectrum in the source rest frame is
\beq
E_{\gamma,c}=\Gamma\frac{3h}{4\pi}\Big(\frac{E_{e,c}^{\prime}}{m_ec^2}\Big)^2
\frac{eB^{\prime}c}{m_ec^2}\simeq 1.9\times10^{-3}~{\rm eV}\Gamma_2 
\Big(\frac{t_{v,-2}\Gamma_{2}^5
f_{c,2}}{L_{iso,51}\epsilon_B({1+Y_e})}\Big)^2B_{5}^{\prime}=2.8{\rm eV}t_{v,-2}
\frac{\xi_{1}^{1/2}}{(L_{iso,51}\epsilon_{B,-1})^{3/2}}\Big(\frac{{\Gamma}_{2}^{4}
f_{c,2}}{1+Y_e}\Big)^{2}
\eeq
Notice that $E_{\gamma,c}$ very sensitively depends on $\Gamma$ and some other parameters
so that it could become a large value when parameters change. For example,  
for $B^{\prime}=10^4$G, $\Gamma=400$, $f_c=500$, $L_{iso}=10^{51}{\rm erg~s^{-1}}$, 
$t_v=0.01$s and $\epsilon_B=0.1$ we get $E_{\gamma,c}\sim 1.9$ MeV. 
The break energy in the photon spectrum due to the minimum electron injection energy is 
\beq
E_{\gamma,m}=\Gamma\frac{3h}{4\pi}\Big(\frac{E_{e,m}^{\prime}}{m_ec^2}\Big)^2
\frac{eB^{\prime}c}{m_ec^2} \simeq 0.58~{\rm MeV} \Gamma_2\Big(\frac{\epsilon_e}
{\epsilon_p}\Big)^2B_5^{\prime}=8.5{\rm MeV}\Big(\frac{\epsilon_e}{\epsilon_p}\Big)^2
(\xi_{1}\epsilon_{B,-1}L_{iso,51})^{1/2}({\Gamma}_{2}^{2} t_{v,-2})^{-1}
\eeq
Assuming $E_{ssa}<E_{\gamma,m,s}<E_{\gamma,c,s}$ the photon energy spectrum from 
synchrotron radiation of slow-cooling relativistic electrons is as follows
\beq 
E_{\gamma,s}^2\frac{dN_{\gamma,s}(E_{\gamma,s})}{dE_{\gamma,s}}\propto\left\{ 
\begin{array}{l@{\quad \quad\quad}l}E_{\gamma,s}^{4/3}
& E_{ssa}<E_{\gamma,s}\leq E_{\gamma,m,s}\\
E_{\gamma,m,s}^{4/3+(p-3)/2}{E_{\gamma,s}^{-(p-3)/2}} &
E_{\gamma,m,s}<E_{\gamma,s}\leq E_{\gamma,c,s}\\
E_{\gamma,m,s}^{4/3+(p-3)/2}E_{\gamma,c,s}^{1/2}E_{\gamma,s}^{-(p-2)/2} 
& E_{\gamma,c,s}\leq E_{\gamma,s}
\end{array}\right.
\label{esynch_spec_slow}
\eeq
In the case of slow-cooling electrons for very small values of $\epsilon_e$ (e.g. $\sim 
10^{-3}$, which is relevant when the hadronic emission component becomes important),
the break in the photon spectrum due to the minimum injection energy of electrons goes 
below the synchrotron self absorption energy. The order in the spectral break energies 
becomes $E_{\gamma,m,s}<E_{ssa}<E_{\gamma,c,s}$, and the spectrum is also modified. The 
spectral indices of the electron synchrotron spectrum for different ordering of the 
spectral break energies are derived by \citet{granot}.
For $E_{\gamma,m,s} < E_{\gamma,s} < E_{ssa}$ the spectral index of 
$E_{\gamma,s}^2\frac{dN_{\gamma,s}(E_{\gamma,s})}{dE_{\gamma,s}}$ is $7/2$, and 
for $E_{\gamma,s} < E_{\gamma,m,s}$ the spectral index is $3$. The 
indices of the spectrum between $E_{ssa}$, $E_{\gamma,c,s}$ and above $E_{\gamma,c,s}$ 
remain as $-(p-3)/2$ and $-(p-2)/2$, respectively.
When $E_{ssa}$ is greater than both $E_{\gamma,m,s}$ and $E_{\gamma,c,s}$ their relative 
ordering becomes unimportant. In that case the spectral indices of 
$E_{\gamma,s}^2\frac{dN_{\gamma,s}(E_{\gamma,s})}{dE_{\gamma,s}}$ are $7/2$ between 
$E_{\gamma,m,s}$ and $E_{ssa}$, and $-(p-2)/2$ above $E_{ssa}$. Below $E_{\gamma,m,s}$ 
the index is $3$.
 
For fast-cooling electrons the synchrotron photon energy spectrum for $E_{ssa}<E_{\gamma,c,s}
<E_{\gamma,m,s}$ is
\beq 
E_{\gamma,s}^2\frac{dN_{\gamma,s}(E_{\gamma,s})}{dE_{\gamma,s}}\propto
\left\{ \begin{array}{l@{\quad \quad\quad}l}E_{\gamma,s}^{4/3} & 
E_{ssa}<E_{\gamma,s}\leq E_{\gamma,c,s}\\
E_{\gamma,c,s}^{5/6}{E_{\gamma,s}^{1/2}} & E_{\gamma,c,s}<E_{\gamma,s}\leq E_{\gamma,m,s}\\
E_{\gamma,c,s}^{5/6}E_{\gamma,m,s}^{(p-1)/2}E_{\gamma,s}^{-(p-2)/2} 
& E_{\gamma,m,s}\leq E_{\gamma,s}
\end{array}\right.
\label{esynch_spec_fast}
\eeq
When the ordering of break energies in the photon spectrum becomes 
$E_{\gamma,c,s}<E_{ssa}<E_{\gamma,m,s}$ the photon energy spectrum is
\beq 
E_{\gamma,s}^2\frac{dN_{\gamma,s}(E_{\gamma,s})}{dE_{\gamma,s}}\propto
\left\{ \begin{array}{l@{\quad \quad\quad}l}E_{\gamma,s}^{13/8} & 
E_{ssa}<E_{\gamma,s}\leq E_{\gamma,c,s}\\
E_{\gamma,c,s}^{9/8}{E_{\gamma,s}^{1/2}} & E_{\gamma,c,s}<E_{\gamma,s}\leq E_{\gamma,m,s}\\
E_{\gamma,c,s}^{9/8}E_{\gamma,m,s}^{(p-1)/2}E_{\gamma,s}^{-(p-2)/2} 
& E_{\gamma,m,s}\leq E_{\gamma,s}
\end{array}\right.
\label{esynch_spec_fast}
\eeq
The total energy emitted in synchrotron radiation by relativistic electrons is 
$E_{iso}\eta_e\epsilon_e/(1+Y_e)$. The normalisation constant for the synchrotron photon 
energy spectrum can be calculated from
\beq
\int_{E_{\gamma,min}}^{E_{\gamma,max}}E_{\gamma,s}\frac{dN_{\gamma,s}(E_{\gamma,s})}
{dE_{\gamma,s}}dE_{\gamma,s}=E_{iso}\frac{\eta_e\epsilon_e}{(1+Y_e)}
\label{norm_synch}
\eeq
The maximum electron energy $E_{e,max}$ can be calculated by equating 
the acceleration time and the shorter of the dynamical and cooling time scales of the 
relativistic electrons.
The expression of the acceleration time scale is 
$t_{acc}^{\prime}=2\pi \zeta r_L(E_{e}^{\prime})/c=
2\pi \zeta E_{e}^{\prime}/eB^{\prime}c$. Here $r_L(E_{e}^{\prime})$ is the Larmor radius of 
an electron of energy $E_{e}^{\prime}$ in a magnetic field $B^{\prime}$,
$\zeta$ can be expressed as $\zeta\sim {\beta_{sh}}^{-2}y$, where  $\beta_{sh}$ is the 
velocity of the shock in the comoving frame of the unshocked medium and $y$ is the 
ratio of diffusion coefficient to the Bohm coefficient \citet{rachen}.  In ultra-relativistic
shocks $\beta_{sh}\approx 1$ and numerical simulations for both parallel and oblique shocks 
gives $\zeta\sim 1$.  With
\beq
t_{acc}^{\prime}=min[t_{cool}^{\prime},t'_{dyn}]~,
\label{emax}
\eeq  
one can derive the maximum comoving electron energy
\beq
E'_{e,max}=min\Big[8.5\Big(\frac{B_{5}^{\prime}\Gamma_{2}^6t_{v,-2}^2}
{L_{iso,51}\epsilon_B({1+Y_e})}\Big)^{1/2},
14.3\times10^7\Gamma_{2}t_{v,-2}B_{5}^{\prime}\Big]{\rm GeV}~
\eeq
For electrons, the cooling term (first term in the bracket) always defines the maximum
electron energy. The maximum synchrotron photon energy in the source rest frame can be 
then derived as 
\beq
E_{\gamma,max}=\Gamma\frac{3h}{4\pi}\Big(\frac{E_{e,max}^{\prime}}{m_ec^2}\Big)^2
\frac{eB^{\prime}c}{m_ec^2} =0.48{\rm GeV}~\Big(\frac{\Gamma_2^7 {B_{5}^{\prime}}^2 
t_{v,-2}^2}{L_{iso,51}\epsilon_{B,-1}(1+Y_e)}\Big)=102{\rm GeV}~\Big(\frac{
{\Gamma}_{2}}{1+Y_e}\Big)
\eeq
This is used in eqn.(\ref{norm_synch}) to define the normalization of the spectrum. 
The result has a very steep dependence on $\Gamma$. We also notice that $B'$ is not
an independent parameter, but can be calculated from other parameters according to
eqn.(\ref{B'}). For example,
for $\Gamma=400$, $L_{iso}=10^{51}$erg/s, $t_v=0.01$s and $\epsilon_B,\epsilon_e
\sim0.1$, the magnetic field is of the order of $10^4$G and the maximum photon 
energy becomes a few hundred GeV. 

\section[]{Electron Inverse Compton Scattering}
The relativistic electrons can be inverse Compton scattered by low energy synchrotron 
photons inside the GRB fireball and transfer their energy to high energy photons. 
Below, we derive the IC photon spectrum using the electron and synchrotron photon 
spectra.
\beq
\frac{dN_{\gamma,i}(E_{\gamma,i})}{dE_{\gamma,i}}\propto \frac{1}{E_{\gamma,i}}\int 
\frac{dN_e(E_e)}{dE_e} dE_e\times\nonumber\int \frac{dN_{\gamma,s}(E_{\gamma,s})}
{dE_{\gamma,s}} dE_{\gamma,s}
\eeq
The electron Lorentz factor (${\gamma_{e}}^{\prime}$), IC and synchrotron photon energies 
($E_{\gamma,i}$, $E_{\gamma,s}$) are related as $E_{\gamma,i}\sim {\gamma_{e}^{\prime}}^2 
E_{\gamma,s}$, this can be used to simplify the above equation. The final expression for 
the IC photon spectrum considering slow cooling of electrons is
\beq 
E_{\gamma,i}^2\frac{dN_{\gamma,i}(E_{\gamma,i})}{dE_{\gamma,i}}\propto\left\{ 
\begin{array}{l@{\quad \quad\quad\quad}l}E_{\gamma,i}^{4/3} 
& \nonumber E_{ssa,i}< E_{\gamma,i}\leq E_{\gamma,m,i}\\
E_{\gamma,m,i}^{4/3+(p-3)/2}{E_{\gamma,i}^{-(p-3)/2}} &
\nonumber E_{\gamma,m,i}<E_{\gamma,i}\leq E_{\gamma,c,i}\\ 
\nonumber E_{\gamma,m,i}^{4/3+(p-3)/2}E_{\gamma,c,i}^{1/2}E_{\gamma,i}^{-(p-2)/2} 
& \nonumber E_{\gamma,c,i}< E_{\gamma,i} \leq E_{\gamma,K}\\ 
\nonumber E_{\gamma,m,i}^{4/3+(p-3)/2}E_{\gamma,c,i}^{1/2}
E_{\gamma,K}^{(p-2)/2}E_{\gamma,i}^{-(p-2)} & \nonumber E_{\gamma,K}< E_{\gamma,i} 
\end{array}\right.
\label{ic_spec_slow}
\eeq
Here $E_{ssa,i}={\gamma_{e,m}^{\prime}}^2E_{ssa}$, $E_{\gamma,m,i}={\gamma_{e,m}^{\prime}}^2
E_{\gamma,m,s}$, and $E_{\gamma,c,i}={\gamma_{e,c}^{\prime}}^2E_{\gamma,c,s}$, 
where ${\gamma_{e,m}^{\prime}}=E'_{e,m}/m_ec^2 = g(p) (m_p/m_e) (\epsilon_e/\epsilon_p)$, 
$\gamma'_{e,c}=E'_{e,c}/m_ec^2$
are Lorentz factors corresponding to the minimum injection energy of electrons and
the cooling break energy of electrons. 
In the case of fast cooling $E_{\gamma,m,i}>E_{\gamma,c,i}$ and the IC photon spectrum has 
to be modified accordingly. 
\beq 
E_{\gamma,i}^2\frac{dN_{\gamma,i}(E_{\gamma,i})}{dE_{\gamma,i}}\propto\left\{ 
\begin{array}{l@{\quad \quad\quad\quad}l}E_{\gamma,i}^{4/3} & 
\nonumber E_{ssa,i}< E_{\gamma,i}\leq E_{\gamma,c,i}\\E_{\gamma,c,i}^{5/6}
{E_{\gamma,i}^{1/2}} &\nonumber
E_{\gamma,c,i}<E_{\gamma,i}\leq E_{\gamma,m,i}\\ \nonumber 
E_{\gamma,c,i}^{5/6}E_{\gamma,m,i}^{(p-1)/2}E_{\gamma,i}^{-(p-2)/2} &
\nonumber E_{\gamma,m,i}< E_{\gamma,i} \leq E_{\gamma,K}\\ 
\nonumber E_{\gamma,c,i}^{5/6}E_{\gamma,m,i}^{(p-1)/2}
E_{\gamma,K}^{(p-2)/2}E_{\gamma,i}^{-(p-2)},&\nonumber E_{\gamma,K}< E_{\gamma,i} 
\end{array}\right.
\label{ic_spec_fast}
\eeq
In eqn.(\ref{ic_spec_fast}) the expressions for $E_{ssa,i}$, $E_{\gamma,c,i}$ and 
$E_{\gamma,m,i}$ are $E_{ssa,i}={\gamma_{e,c}^{\prime}}^2E_{ssa}$, 
$E_{\gamma,c,i}={\gamma_{e,c}^{\prime}}^2E_{\gamma,c,s}$ and 
$E_{\gamma,m,i}={\gamma_{e,m}^{\prime}}^2E_{\gamma,m,s}$. 
When $E_eE_{\gamma,s}>>\Gamma^2 m_e^2c^4$ the cross section for IC scattering decreases 
as the scattering enters the Klein Nishina (KN) regime. A break in the photon spectrum at 
$E_{\gamma,i}=E_{\gamma,K}$ appears when the Klein Nishina effect becomes important. We 
define a parameter $\kappa=\frac{E_eE_{\gamma,peak}}{\Gamma^2m_e^2c^4}$, where 
$E_{\gamma,peak}=max[E_{\gamma,c,s};E_{\gamma,m,s}]$. The KN regime starts when 
$\kappa=1$ (e.g. Fragile et al. 2004), and
\beq
E_{\gamma,K}=\frac{\Gamma^2m_e^2c^4}{E_{\gamma,peak}}=2.5~{\rm GeV}\frac{\Gamma_{2}^2}
{E_{\gamma,peak,MeV}}
\eeq
In the KN regime the emissivity of electrons decreases by $\kappa^2$, and the
photon energy spectral index simply follows the electron energy spectral index,
i.e. $-(p-2)$. The IC photon spectrum in eqn.(\ref{ic_spec_slow}) can be normalised 
as 
\beq
\int_{E_{\gamma,m,i}}^{E_{\gamma,max,i}} E_{\gamma,i}\frac{dN_{\gamma,i}
(E_{\gamma,i})}{dE_{\gamma,i}}dE_{\gamma,i}=E_{iso}\frac{\eta_e\epsilon_e Y_e}{1+Y_e}~, 
\eeq
where $E_{\gamma,max,i}=\Gamma E'_{e,max}$ due to the KN effect.

\section[]{Proton Synchrotron Radiation}
Relativistic protons lose energy by synchrotron radiation and photo-pion 
($\pi^0$, $\pi^{+}$) production inside GRBs. 
They interact with the low energy photons in the GRB environment and pions are produced. 
There is a threshold energy for this interaction ($p\gamma$) to happen, $E_p E_{\gamma}
\geq 0.3GeV^2\Gamma^2$, where $E_p$ and $E_{\gamma}$ are proton, photon energy in the 
source rest frame respectively. The $\pi^0$s decay to a pair of high energy photons, 
while the $\pi^+$s decay to neutrinos and leptons. 
The threshold condition therefore suggests that the photon-pion related high energy
spectrum is typically more energetic than the electron IC spectrum.
We assume that the proton spectrum in the internal shocks 
can be expressed as a power law in proton energy. We 
consider a proton spectral index similar to electrons for our present discussion. 
Since protons are poor emitters, we only consider the 
scenario of slow-cooling in the comoving proton spectrum 
\beq
\frac{dN_p(E'_p)}{dE'_p}\propto\left\{ \begin{array}{l@{\quad \quad}l}
{{E'_p}^{-p}} & E'_{p,m}<E'_p<E'_{p,c}\\{E'_p}^{-p-1} & E'_{p,c}<E'_p
\end{array}\right.
\label{proton_spec}
\eeq
where $E'_{p,m}$ is the minimum injection energy of the protons and $E'_{p,c}$ is break 
energy in the spectrum due to proton cooling.
The minimum injection energy $E'_{p,m}=\bar{\gamma'_p} m_pc^2g(p)$, where 
$g(p)=\frac{p-2}{p-1}$ for $p \gg 2$ and $g(p)\sim 1/6$ for $p=2$.
The cooling break energy can be derived by comparing the comoving and the cooling time 
scales. The inverse of the cooling time scale $t_{cool}^{\prime}$ of a proton is
\beq
\frac{1}{t_{cool}^{\prime}}=\frac{1}{t_{syn}^{\prime}}+\frac{1}{t_{\pi}^{\prime}}
\label{p_cool}
\eeq
The photo-pion cooling time scale $t'_\pi$ has been derived earlier in the context of 
estimation of neutrino fluxes from GRBs \citep{wax,nayan}. If $f_{\pi}$ is the fraction 
of proton energy going to pion production in the $\Delta$ resonance of $p\gamma$ 
interactions one has $1/t_{\pi}^{\prime} \sim f_{\pi}/t'_{dyn}$ where the comoving time 
scale is\footnote{In this definition, on average protons loose 
$\sim 20\%$ energy in the time scale of $t'_\pi$. Although it is not strictly the 
e-folding timescale usually used to define cooling, for order-of-magnitude estimates
this is good enough.} $t'_{dyn}=\Gamma t_v$. 
The peak value of $p\gamma$ interaction cross section at the 
$\Delta$ resonance is $\sigma_{p\gamma}=5\times10^{-28}{\rm cm}^2$. 
This is much higher than the Thomson cross section for protons $\sigma_{p,T}=
\Big(\frac{m_e}{m_p}\Big)^2 \sigma_{e,T}$, where $\sigma_{e,T}=6.625\times10^{-25}
{\rm cm}^2$. We therefore neglect the IC process of protons.
Substituting for $t_{syn}^{\prime}$ and $t_{\pi}^{\prime}$ in eqn.(\ref{p_cool}), we get
\beq
\frac{1}{t_{cool}^{\prime}}=\frac{4}{3}\sigma_{p,T}{\beta_{p}^{\prime}}^2\frac{E_p^{\prime}}
{m_pc^2}\frac{cU\epsilon_B}{m_pc^2}+\frac{f_{\pi}}{\Gamma t_v}
\eeq
where, $\beta_{p}^{\prime}$ is dimensionless speed of relativistic protons.
 We use the general expression for $f_{\pi}$ from \citet{nayan}
\beq
f_{\pi}(E_p) =f_{0} \left\{\begin{array}{l@{\quad \quad}l}
\frac{1.34^{\alpha_2-1}}{\alpha_2+1}(\frac{E_{p}}{E_{pb}})^{\alpha_2-1}
& E_p<E_{pb}\\\frac{1.34^{\alpha_1-1}}{\alpha_1+1}
(\frac{E_{p}}{E_{pb}})^{\alpha_1-1} 
& E_{p}>E_{pb}\end{array} \right.
\label{fpi}
\eeq 
where 
\beq
f_{0}=\frac{0.9
L_{iso,51}}{810\Gamma_{2}^4t_{v,-2}E_{\gamma,peak,MeV}}\frac{1}
{[\frac{1}{\alpha_2-2}-\frac{1}{\alpha_1-2}]}\frac{\eta_e \epsilon_e}{1+Y_e}~. 
\label{f0}
\eeq
In our 
present discussion $\alpha_2=(p+2)/2$  and $\alpha_1=(p+1)/2$.
$E_{\gamma,peak,MeV}$ is the peak energy in the electron synchrotron photon spectrum 
expressed in MeV, and $L_{iso,51}$ is the GRB luminosity in unit of $10^{51}
~{\rm erg~s^{-1}}$, which is the typical value for GRB luminosities. $E_{pb}=0.3\Gamma^2
/E_{\gamma,peak,GeV}$GeV is the threshold proton energy for interaction with photons of 
energy $E_{\gamma,peak,GeV}$. For typically observed values of GRB parameters one 
has $E_{pb}\sim 1$ PeV. The break energy in the proton spectrum due to proton cooling 
can be calculated by comparing the comoving and cooling time scales of protons as 
discussed in the case of electrons in \S2. We assume $\beta_{p}^{\prime}\sim 1$ then for 
$E_p<E_{pb}$ the expression of cooling break energy in the comoving frame is
\beq
E_{p,c}^{\prime}=\frac{f_{c}}{\Gamma t_{v}}\Big(\frac{4}{3}\sigma_{p,T}
{\beta_{p}^{\prime}}^2
\frac{cU\epsilon_B}{m_p^2c^4}+\frac{f_0}{E_{pb}\Gamma t_{v}}\frac{1.34^{\alpha_2-1}}
{\alpha_2+1}\Big)^{-1}=\frac{10^{8}
{\rm GeV} f_{c,2}}{\Gamma_{2}t_{v,-2}}
\Big(0.16\frac{L_{iso,51}\epsilon_B}{{\Gamma_{2}}^6t_{v,-2}^2}+\frac{f_0}{
E_{pb}({\rm PeV})\Gamma_2t_{v,-2}}\frac{1.34^{\alpha_2-1}}{\alpha_2+1}\Big)^{-1}
\eeq
 where $f_c=\frac{t'_{dyn}}{t_{cool}^{\prime}}$.
 The synchrotron photon spectrum from relativistic protons is
\beq
E_{\gamma,ps}^2\frac{dN_{\gamma,ps}(E_{\gamma,ps})}{dE_{\gamma,ps}}
\propto\left\{ \begin{array}{l@{\quad\quad}l}{E_{\gamma,ps}^{-(p-3)/2}}&
E_{\gamma,m,ps}<E_{\gamma,ps}\leq E_{\gamma,c,ps}\\E_{\gamma,c,ps}^{1/2}
E_{\gamma,ps}^{-(p-2)/2} & E_{\gamma,c,ps}< E_{\gamma,ps}
\end{array}\right.
\eeq
The minimum injection energy in the photon spectrum from proton synchrotron radiation 
is related to that from electron synchrotron radiation as \citep{zhang1}
\beq
\frac{E_{\gamma,m,ps}}{E_{\gamma,m,s}}=\Big(\frac{E'_{p,m}}{E'_{e,m}}\Big)^2
\Big(\frac{m_e}{m_p}\Big)^3
\eeq
The cooling break energy in the photon spectrum from proton synchrotron radiation is the 
characteristic synchrotron photon energy for proton energy $E'_{p,c}$. To normalize the
proton synchrotron spectrum, it is important to find out the relative importance 
between proton synchrotron radiation and $p\gamma$ interactions. Similar to the 
treatment of electrons, one can define
\beq
Y_p = \frac{L_{p,p\gamma}}{L_{p,syn}} = \frac{\sigma_{p\gamma}}{\sigma_{p,T}}
\frac{U_{e,syn}}{U_B} =  \frac{\sigma_{p\gamma}}{\sigma_{p,T}} Y_e~.
\label{Yp}
\eeq
where, $L_{p,p\gamma}$ and $L_{p,syn}$ are the luminosities of radiations emitted in $p\gamma$ interactions and synchrotron emission of protons respectively. 
Notice that protons interact with the synchrotron emission of the electrons, so
that $Y_e$ enters the problem. Eqn. (\ref{Yp}) suggests that $Y_p$ is usually much
greater than unity since $\sigma_{p\gamma} \gg \sigma_{p,T}$. As a result, most of 
the proton energy is lost through $p\gamma$ interaction rather than proton
synchrotron radiation.

The proton synchrotron photon spectrum can be normalised as
\beq
\int_{E_{\gamma,m,ps}}^{E_{\gamma,max,ps}}E_{\gamma,ps}\frac{dN_{\gamma,ps}(E_{\gamma,ps})}
{dE_{\gamma,ps}}dE_{\gamma,ps}=E_{iso}\frac{\epsilon_p\eta_p}{1+Y_p}~, 
\label{psynch_spec}
\eeq
where $\eta_p=
\Big({E'_{p,c}}/{E'_{p,m}}\Big)^{2-\alpha}$. The maximum proton
synchrotron photon energy is derived by $E_{\gamma,max,ps}=\Gamma\frac{3h}{4\pi}
\Big(\frac{E_{p,max}^{\prime}}{m_pc^2}\Big)^2\frac{eB^{\prime}c}{m_pc^2}$, where
$E'_{p,max}$ is again defined by comparing the comoving acceleration time with the shorter 
of the comoving dynamical and cooling times scales 
\beq
E'_{p,max}={\rm min}
\Big[50\Big(\frac{B_{6}^{\prime}\Gamma_{2}^6t_{v,-2}^2}{L_{iso,51}\epsilon_B
({1+Y_p})}\Big)^{1/2}, 1.4\times10^6\Gamma_{2}t_{v,-2}B_{6}^{\prime}\Big]{\rm TeV}~.
\eeq
or,
\beq
E'_{p,max}={\rm min}\Big[191\Big(\Big(\frac{\xi_{1}}{\epsilon_{B,-1}L_{iso,51}}
\Big)^{1/2}\frac{{\Gamma}_{2}^3 t_{v,-2}}{1+Y_p}\Big)^{1/2},\frac{208}
{{\Gamma}_{2}^2}\times10^4(\xi_{1}\epsilon_{B,-1}L_{iso,51})^{1/2}\Big]{\rm TeV}~.
\eeq

\section{$\pi^0$ Decay}
The relativistic protons interact with the low energy photons and
photo-pions ($\pi^0$,$\pi^{+}$) are produced as a result. The probabilities of 
$\pi^0$ and $\pi^{+}$ production are
1/3 and 2/3, respectively. Pions subsequently decay,
i.e. $\pi^{0}\rightarrow \gamma \gamma$ and $\pi^{+}\rightarrow
\mu^{+}\nu_{\mu}\rightarrow \nu_{\mu}\bar\nu_{\mu}\nu_e e^+$.  As the
cross section for the $\gamma\gamma$ interactions is much higher than
the peak value of $p\gamma$ interaction cross section, above the
threshold energy of pair production $\gamma\gamma$ interactions are
expected to dominate over $p\gamma$ interactions. If the photon energy
is $2m_ec^2 \sim 1$ MeV in the comoving frame, then in the source rest
frame it is of the order of a few hundred MeV as the Lorentz factors
are typically of the order of few hundred for canonical GRBs. For
example, for $\Gamma=400$ the photons of energy 400 MeV can produce
photo-pions by interaction with protons of minimum energy $E_p\sim
120$ TeV. The $\pi^0$ typically carries $20\%$ of the proton's energy
and the photons produced in $\pi^0$ decay share its energy
equally. Hence, the minimum energy of the photons produced from
$\pi^0$ decay is expected to be $\sim 10\% E_p \sim 12$ TeV. The
photon spectrum produced from $\pi^0$ decay has been derived below
using the proton spectrum defined in eqn.(\ref{proton_spec}) and 
assuming the fraction $f_{\pi}/3$ of protons' energy goes to $\pi^0$s. 
\beq
E_{\gamma,\pi^0}^2\frac{dN_{\gamma,\pi^0}(E_{\gamma,\pi^0})}{dE_{\gamma,\pi^0}}
\propto\frac{1}{3}\frac{f_{\pi}(E_{\gamma,\pi^0})}{2}\left\{\begin{array}
{l@{\quad\quad}l}E_{\gamma,\pi^0}^{2-p} & E_{\gamma,\pi^0} \leq E_{\gamma,\pi^0,c}
\\E_{\gamma,\pi^0}^{1-p} & E_{\gamma,\pi^0}>E_{\gamma,\pi^0,c}
\end{array}\right.
\eeq
where, $E_{\gamma,\pi^0,c}=0.1E_{p,c}$.  For the expression for
$f_{\pi}$, see eqn.({\ref{fpi}}), which contains a break energy. The
break energy in the photon spectrum contained within $f_{\pi}$ is
$E_{\gamma,\pi^0,b}= 0.03\Gamma^2/\epsilon_{br,GeV}$ GeV assuming
$10\%$ of the proton's energy goes to the photon produced via $\pi^0$
decay. $\epsilon_{br}$ is the break energy in the low energy photon
spectrum (in the scenario of slowly cooling electrons it is the
cooling break energy in the photon spectrum and for fast cooling
electrons it is the photon energy corresponding to the minimum
injection energy of electrons).  The photon flux can be normalised in
the following way
\bea
\int_{E_{\gamma,\pi^0,min}}^{E_{\gamma,\pi^0,max}}E_{\gamma,\pi^0} \frac{dN_{\gamma,\pi^0}
(E_{\gamma,\pi^0})}{dE_{\gamma,\pi^0}}dE_{\gamma,\pi^0}=\frac{E_{iso}}{3}
\frac{\epsilon_p\eta_p Y_p}
{1+Y_p}
\label{photo-pion_spec}
\eea 
where $E_{\gamma,\pi^0,min}=30\Gamma$ GeV and $E_{\gamma,\pi^{0},max}=0.1E_{p,max}$. 
Although high energy photons ($\sim$ TeV) are absorbed by lower energy photons and 
$e^+e^-$ pairs are produced, at extreme energies the pair production cross 
section decreases with increasing energy \citep{raz1}. Hence, ultrahigh energy photons can 
escape from the internal shocks for suitable parameters depending on the values of their 
various parameters and the low energy photon spectra.
 
\section{Synchrotron Radiation 
of Positrons Produced in $\pi^+$ decay}
The shock accelerated protons may interact with the low energy photons
to produce $\pi^+$s along with $\pi^0$s as discussed in the previous
section. The $\pi^{+}$s subsequently decay to muons and neutrinos. The
energetic muons decay to positrons and neutrinos ($p\gamma\rightarrow
\pi^+\rightarrow \mu^{+}\nu_{\mu}
\rightarrow e^+\nu_{\mu}\bar\nu_{\mu}\nu_e$). The charged pions, 
muons and the positrons are 
expected to lose energy through synchrotron radiation and IC inside
the shock region.  As the Thomson cross section for positrons is much
larger than pions or muons, they are expected to emit much more
radiation compared to the heavier charged particles. On the other
hand, since these positrons are very energetic, most IC processes
happen in the Klein Nishina regime. We therefore neglect the
contribution of the positron IC processes.  The positron synchrotron
spectrum produced in $p\gamma$ interactions can be derived in the
following way. The fraction of the protons' energy tranferred to pions
is denoted by $f_{\pi}$ (eqn[\ref{fpi}]). If we assume that the final state
leptons share the pion's energy equally then one fourth of the pion's
energy goes to the positron. The energy of the positron spectrum
$\frac{dN(E_{e^{+}})}{dE_{e^+}}$ at the energy $E_{e^{+}}$ can be
expressed using the proton spectrum defined in eqn.(\ref{proton_spec})
\beq
E_{e^{+}}^2\frac{dN(E_{e^{+}})}{dE_{e^{+}}}\propto\frac{2}{3}\frac{f_{\pi}
(E_{e^{+}})}{4}\left\{\begin{array}{l@{\quad\quad}l}E_{e^{+}}^{2-p} & E_{e^{+}} 
\leq E_{e^{+},c}
\\E_{e^{+}}^{1-p} & E_{e^{+}}>E_{e^{+},c}
\end{array}\right.
\label{pos_spec}
\eeq
where, $E_{e^{+},c}$ is the cooling break energy in the positron specrum and
\beq
f_{\pi}(E_{e^+}) =f_{0} \left\{\begin{array}{l@{\quad \quad}l}
\frac{1.34^{\alpha_2-1}}{\alpha_2+1}(\frac{E_{e^+}}{E_{e^{+}b}})^{\alpha_2-1}
& E_{e^+}<E_{e^+b}\\\frac{1.34^{\alpha_1-1}}{\alpha_1+1}
(\frac{E_{e^+}}{E_{e^{+}b}})^{\alpha_1-1} 
& E_{e^{+}}>E_{e^{+}b}\end{array} \right.
\label{fpi_e^+}
\eeq 
where $f_0$ has been defined in eqn.(\ref{f0}), $E_{e^+b}=0.05E_{pb}$, 
$E_{pb}=0.3~{\rm GeV}\Gamma^2/\epsilon_{br,GeV}$, and $\epsilon_{br,GeV}$ 
is the break energy in the photon spectrum as defined earlier. 
The positron spectrum in eqn.(\ref{pos_spec}) can be normalised using the 
total energy carried by the positrons,
\beq
\int_{E_{e^{+},min}}^{E_{e^{+},max}}E_{e^+}\frac{dN(E_{e^+})(E_{e^+})}
{dE_{e^+}}dE_{e^+}=\frac{1}{6}\frac{\epsilon_p\eta_pY_pE_{iso}}{1+Y_p}
\eeq
The maximum and minimum positron energies are $E_{e^{+},max}=0.05E_{p,max}$
 and $E_{e^{+},min}=15\Gamma$ GeV (which is $\sim 6$TeV for $\Gamma=400$).
The synchrotron photon spectrum from the positrons  
can be subsequently derived using the same treatment for primary electrons 
as discussed in \S2. The IC emission is in the KN regime and therefore not
important. Also, photons having energies above a few hundred GeV are annihilated by lower energy photons as discussed in the following section.
The relativistic muons produced in $\pi^+$ decay lose energy by synchrotron radiation. We compare the decay and synchrotron energy loss time scales of the high energy muons. The maximum energies of positrons can be calculated in this way. If the muons decay before losing energy significantly high energy positrons are produced carrying approximately $5\%$ of the initial proton's energy. On the otherhand if the muons lose energy before they decay lower energy positrons are produced. These positrons radiate energy and produce lower energy photons. The 
muons initially carry approximately $10\%$ of the relativistic protons' energy hence, we expect the low energy photon flux produced by cooling of positrons is lower than that produced by relativistic electrons if $\epsilon_e$ and $\epsilon_p$ are comparable.   
\section[]{Internal Pair-Production Optical Depths of High Energy Photons}
Inside GRBs high energy photons interact with low energy photons to produce 
electron-positron pairs (e.g. Baring \& Harding 1997; Lithwick \& Sari 2001). 
The optical depth depends on the values of various
parameters of the GRB fireball. We follow the approach discussed in \citet{bhat} 
to derive internal optical depths of GRBs in detail. For two photons (a high energy
photon $\gamma_h$ and a low energy photon $\gamma_l$), the pair production cross 
section depends on the energies of the photons and the angle between their 
directions of propagation. The cross section is 
\citep{ber}
\bea
\sigma_{\gamma_h\gamma_l}(E_{\gamma_h}^{\prime},E_{\gamma_l}^{\prime},\theta)=
\frac{3}{16}\sigma_T(1-{\beta^{\prime}}^2)\Big[(3-{\beta^{\prime}}^4)\ln{\frac
{1+{\beta^{\prime}}}{1-{\beta^{\prime}}}}-2{\beta^{\prime}}(2-{\beta^{\prime}}^2)\Big]
\label{cross_sec}
\eea
where $\sigma_T$ is the Thomson cross section, and $\beta^{\prime}=
[1-(E_{\gamma_l,th}^{\prime}/E_{\gamma_l}^{\prime})]^{1/2}$ is the center of mass 
dimensionless speed of the pair produced. The threshold energy of pair 
production with a high energy photon of energy $E_{\gamma_h}^{\prime}$ is 
\beq
E_{\gamma_l,th}^{\prime}=\frac{2(m_ec^2)^2}{E_{\gamma_h}^{\prime}(1-cos{\theta})}
\eeq
For the photons with energy higher than the threshold energy, the pair 
production cross section decreases with 
increasing photon energy \citep{jau,raz1}. In the present work we calculate
internal optical depths in different energy regimes using the cross sections with 
different energy dependences. The mean free path for $\gamma_h$ $\gamma_l$ interactions 
$l_{\gamma_h\gamma_l}$ can be calculated using the low energy photon spectrum.
\bea
l_{\gamma_h\gamma_l\theta}^{-1}(E_{\gamma_h}^{\prime},\theta)=
\int_{E_{\gamma_l,th}^{\prime}}^{\infty} d{E_{\gamma_l}^{\prime}}
\frac{dn_{\gamma_l}(E_{\gamma_l}^{\prime})}{d{E_{\gamma_l}^{\prime}}}
\sigma_{\gamma_h\gamma_l}(E_{\gamma_h}^{\prime},E_{\gamma_l}^{\prime},\theta)
\eea
and,
\beq
l_{\gamma_h\gamma_l}^{-1}(E_{\gamma_h}^{\prime})=\frac{1}{2}\int_{-1}^{+1} 
d(cos{\theta})(1-cos{\theta})l_{\gamma_h\gamma_l\theta}^{-1}(E_{\gamma_h}^{\prime},\theta)
\label{path_l}
\eeq
where $\frac{dn_{\gamma_l}(E_{\gamma_l}^{\prime})}{dE_{\gamma_l}^{\prime}}$ is the 
specific number density of 
low energy photons inside the GRB. The low energy photon spectrum is observationally 
known, as revealed by gamma-ray detectors such as BATSE and Swift. Theoretically, it 
corresponds to the electron synchrotron component as discussed in \S2, which is a 
broken power law spectrum separated by the synchrotron self absorption break, the 
minimum injection break and the cooling break. The low energy photon flux is related 
to the observed luminosity through
\beq
\int_{E_{\gamma_l,ssa}^{\prime}}^{E_{\gamma_l,max}^{\prime}}{E_{\gamma_l}}^{\prime}
\frac{dn_{\gamma_l}(E_{\gamma_l}^{\prime})}
{dE_{\gamma_l}^{\prime}}{dE_{\gamma_l}}^{\prime}=U_{\gamma}=\frac{L_{\gamma,iso}}
{4\pi c {r_{is}}^2 \Gamma^2}
\eeq 
where $L_{\gamma,iso}$ is the isotropic $\gamma$-ray luminosity. We have taken it to 
be equal to the luminosity of the synchrotron photons emitted by electrons: 
$L_{\gamma,iso}=L_{e,syn}=\frac{\epsilon_e\eta_eL_{iso}}{1+Y_e}$. 
In eqn.(\ref{path_l}) we have three variables: angle $\theta$ and photon energies 
$E_{\gamma_l}^{\prime}$, $E_{\gamma_h}^{\prime}$. To simplify the integration in 
eqn.(\ref{path_l}) we transform the integral with a new variable following \citet{gold}
\beq
s=\frac{E_{\gamma_l}^{\prime}E_{\gamma_h}^{\prime}(1-cos{\theta})}{2(m_ec^2)^2}=
\frac{E_{\gamma_l}^{\prime}}{E_{\gamma_l,th}^{\prime}}=s_0 \Theta
\eeq
with
$s_0=\frac{E_{\gamma_l}^{\prime}E_{\gamma_h}^{\prime}}{(m_ec^2)^2}$, and $\Theta=
\frac{1}{2}(1-cos{\theta})$. As $\beta^{\prime}=(1-1/s)^{1/2}$, the pair production 
cross section can be expressed as a function of the new variable $s$. It is then possible 
to write eqn.(\ref{path_l}) as 
\bea
l_{\gamma_h\gamma_l}^{-1}(E_{\gamma_h}^{\prime})=\frac{3}{8}\sigma_T\Big(\frac{m_e^2c^4}
{E_{\gamma_h}^{\prime}}\Big)^2\int_{\frac{m_e^2c^4}{E_{\gamma_h}^{\prime}}}^{\infty} 
\left[{E_{\gamma_l}^{\prime}}^{-2}\frac{dn_{\gamma_l}(E_{\gamma_l}^{\prime})}{dE_{\gamma_l}
^{\prime}} dE'_{\gamma l} \right] Q[{s_0(E_{\gamma_l}^{\prime})}]
\label{path_l2}
\eea
where
\beq
Q[s_0(E_{\gamma_l}^{\prime})]=\int_{1}^{s_0(E_{\gamma_l}^{\prime})} s\sigma(s)ds~,
\label{del_s0}
\eeq
and $\sigma(s)=\frac{16}{3}\frac{\sigma_{\gamma_h\gamma_l}}{\sigma_{T}}$. 
For moderate values of $s$ we use $\sigma(s)\simeq 1$ and for $s>>1$ it can be 
approximated as $\sigma(s)\simeq\ln(s)/s$. The expressions for 
$Q[s_0(E_{\gamma_l}^{\prime})]$ 
are $(s_0^2-1)/2$ and $s_0(\ln s_0 -1)$, respectively, in the two cases.
Substituting for $Q[s_0(E_{\gamma_l}^{\prime})]$ in eqn.(\ref{path_l2}) we derive the 
final expression for $l_{\gamma_h\gamma_l}^{-1}(E_{\gamma_h}^{\prime})$. The internal 
optical depth $\tau_{int}(E_{\gamma_h}^{\prime})$ is the ratio of comoving time scale and 
the mean time between two pair production interactions.
\beq
\tau_{int}(E_{\gamma_h}^{\prime})=\frac{r_{is}}{\Gamma c}c l_{\gamma_h\gamma_l}^{-1}
(E_{\gamma_h}^{\prime})
\eeq
The final photon energy spectrum to be observed on Earth from nearby GRBs (neglecting
further attenuations with the infrared background and cosmic microwave background)
can be obtained by correcting the original flux for the internal optical depth and 
the redshift $z$ of the source
\beq
E_{\gamma,ob}^2\frac{dN_{\gamma,ob}(E_{\gamma,ob})}{dE_{\gamma,ob}}=
\frac{1}{4\pi d_{z}^2(1+z)}E_{\gamma}^2\frac{dN_{\gamma}(E_{\gamma})}{dE_{\gamma}}
\exp(-\tau_{int}(E_{\gamma}))~,
\label{obs}
\eeq 
where 
\beq
d_z=\frac{c}{H_0}\int_{0}^{z}\frac{dz^{\prime}}{\sqrt{\Omega_{\Lambda}+\Omega_{m}
(1+z^{\prime})^3}}
\eeq
is the comoving distance of the source, 
$H_0=71 {\rm km~ s^{-1}~ Mpc^{-1}}$ is the Hubble constant, and $\Omega_{\Lambda}=0.73$ 
and $\Omega_{m}=0.27$ are adopted in our calculations.

\section[]{Photon Spectrum from Secondary Electrons and Positrons}

The secondary pairs carry a significant fraction of energy in the primary spectrum, 
and this energy is re-radiated and converted to photons. A more realistic treatment
should consider a photon-pair cascade process, which requires numerical calculations
\citep{peer1,peer2}. Here instead we estimate the emission 
from the secondary pairs. We first calculate the photon energy spectra generated by 
different physical processes as discussed earlier. The photon spectra are then corrected 
for internal optical depths and subsequently the total energies carried by these photons 
are calculated by integrating the corrected photon energy spectra over photon energies. 
If we subtract the total energies carried by these high energy photons from their intial 
energies before including the effects of internal optical depths, we get the energies of 
the secondary $e^{-}$ and $e^{+}$ produced in $\gamma\gamma$ interactions. These pairs 
are expected to have spectral indices similar to the high energy photons. With the 
knowledge of their spectral indices and the total energies carried by them the 
synchrotron photon spectra radiated by these secondary leptons are calculated. 
For the parameters adopted in this paper, it turns out that the emission contribution
from the secondaries is below the emission level of the primaries, and hence, does not
significantly modify the observed the spectrum. We therefore do not include this
component in Figs.1-5, but caution that such a feedback process could be potentially
important for the parameter regimes with high opacity. We refer to \citet{peer1} and
\citet{peer2} for more detailed treatments of such cases.

\section[]{Synthesized Spectra and Detectability}

Using the procedure delineated above, we have calculated the broad-band emission 
spectrum from internal shocks for a wide range of parameter regimes. In particular 
we focus on the various high energy emission components discussed above and their 
relative significance. Our results are presented in Fig.1-5. In each set of
calculations we have presented the internal optical depth after the final
photon energy spectrum. For particles accelerated by ultra-relativistic shocks the 
spectral index is expected to be about 2.26 \citet{lemoine}. Afterglow modeling
suggests a larger scatter of $p$ values for relativistic shocks, but $p=2.3$ is
close the mean value of the data (Panaitescu \& Kumar 2002). In all our calculations, 
the spectral indices of relativistic electrons and protons are both assumed as $p=2.3$. 

Figures 1-4 are the calculations for a typical long GRB with duration $T_{90}=20$ s
at redshift $z=1$ (10 s in the source rest frame). Since we
do not know the physical condition of the internal shocks from the first principles, 
we vary the parameter regime in a wide range. In each set of
calculations, we design the parameters to make the electron synchrotron emission
peaking at the sub-MeV range ($\sim$ 0.36 MeV, 0.13 MeV, 0.6 MeV, 0.25 MeV for
Figs.1-4, respectively), as suggested by the data. The global energetics of
the GRB is also adjusted so that the gamma-ray luminosity in the sub-MeV range
is about $10^{51}~{\rm ergs~s^{-1}}$ as suggested by the observations.
The variability time scale
for these calculations is taken as $t_v = 0.01$ s. The bulk Lorentz factor
is adopted as $\Gamma=400$ in Fig.1-3, as suggested by the recent early optical 
afterglow observations (Molinari et al. 2006). In order to check how $\Gamma$ affects
the spectra, we also calculate the case of $\Gamma=1000$ for the parameter set of
Fig.1, which is presented in Fig.4. 
In all the figures, the different components of the photon energy spectrum from a 
GRB for both electrons ($e$) and protons ($p$) are displayed with different line 
styles/colours. The observed energy fluxes $E_{\gamma,ob}^{2}\frac{dN_{\gamma,ob}
(E_{\gamma,ob})} {E_{\gamma,ob}}$ in unit of ergs/cm$^2$sec are plotted against 
the observed photon energy $E_{\gamma,ob}(eV)$. The green long dashed curves 
represent the synchrotron emission from the relativistic electrons. The short 
dashed curves (blue) represent the IC spectrum from energetic electrons; the 
dash-dotted curves (light blue) represents the synchrotron emission of the 
relativistic protons; the triple short dashed curves (orange) represent for the 
synchrotron emission of the relativistic positrons produced in $\pi^{+}$ decays; 
the ultrahigh energy emission component from $\pi^{0}$ decays is shown by the 
double short dashed curves (black) in the extremely high energy regime.
The thin black solid lines represent the synthesized spectra of various components
without including the effect of pair production attenuation. Depending on parameters, 
the pair opacity becomes important in the GeV - TeV range. The thick black solid lines 
represent the final photon spectrum after including the internal optical depths. 
In order to check whether the predicted high energy components are detectable by
GLAST, we also plot an indicative GLAST sensitivity threshold in the 100 MeV - 100 
GeV energy range. The GLAST sensitivity estimate is based on the criterion of 
detecting at least a few photons in the band based on the average effective area
and photon incoming zenith angle of LAT. Background is negligible for GRB detections.
This gives a rough fluence threshold of $\sim 2\times 10^{-7}~{\rm erg~cm^{-2}}$
(B. Dingus, 2007, personal communication). 
The flux thresholds adopted in all the figures are therefore derived from the
observed durations. For $T_{90}=20$ s, this gives a flux threshold of $\sim 
10^{-8}~{\rm erg~cm^{-2}~s^{-1}}$. The sensitivity of VERITAS to photon above 
energy 200GeV has also been shown in our figures with pink dotted line. It is 
$2\times 10^{-8}{\rm erg~cm^{-2}~s^{-1}}$ (D. Horan, 2007, personal communication). 

Figure 1 is a standard ``slow-cooling'' leptonic-dominant case. The shock equipartition 
parameters are $\epsilon_e=0.4$ and $\epsilon_B=0.2$. 
The isotropic shock luminosity is $L_{iso}=10^{52}~{\rm erg~s^{-1}}$. The slow cooling
factor $f_c=2500$ is adopted, which suggests that the post-shock magnetic field
decays on a length scale shorter than the comoving scale (Pe'er \& Zhang 2006).
The thick black line shown on the right side around $10^{15}$eV is the $\pi^0$ 
component after including the effect of absorption due to pair production,
indicating the reduction of pair opacity at high energies (Fig.1b, see also
Razzaque et al. 2004). In this figure the break energies in the photon energy 
spectrum appear in the order of $E_{ssa}<E_{\gamma,m}<E_{\gamma,c}$ in the electron 
synchrotron and IC spectral components. 
The spectral index of the photon energy spectrum is $4/3$ between $E_{ssa}$ and 
$E_{\gamma,m}$, $-(p-3)/2$ between $E_{\gamma,m}$ and $E_{\gamma,c}$, and $-(p-2)/2$ 
above $E_{\gamma,c}$. Since $\epsilon_e$ is large, the leptonic components are
many orders of magnitude stronger than the hadronic components.
The value of $Y_p$ is much larger than 1, so that the proton 
synchrotron component is below the components due to $\pi^0$ decay and positron 
synchrotron radiation. 

\begin{figure}
\begin{center}
\includegraphics[width=10cm]{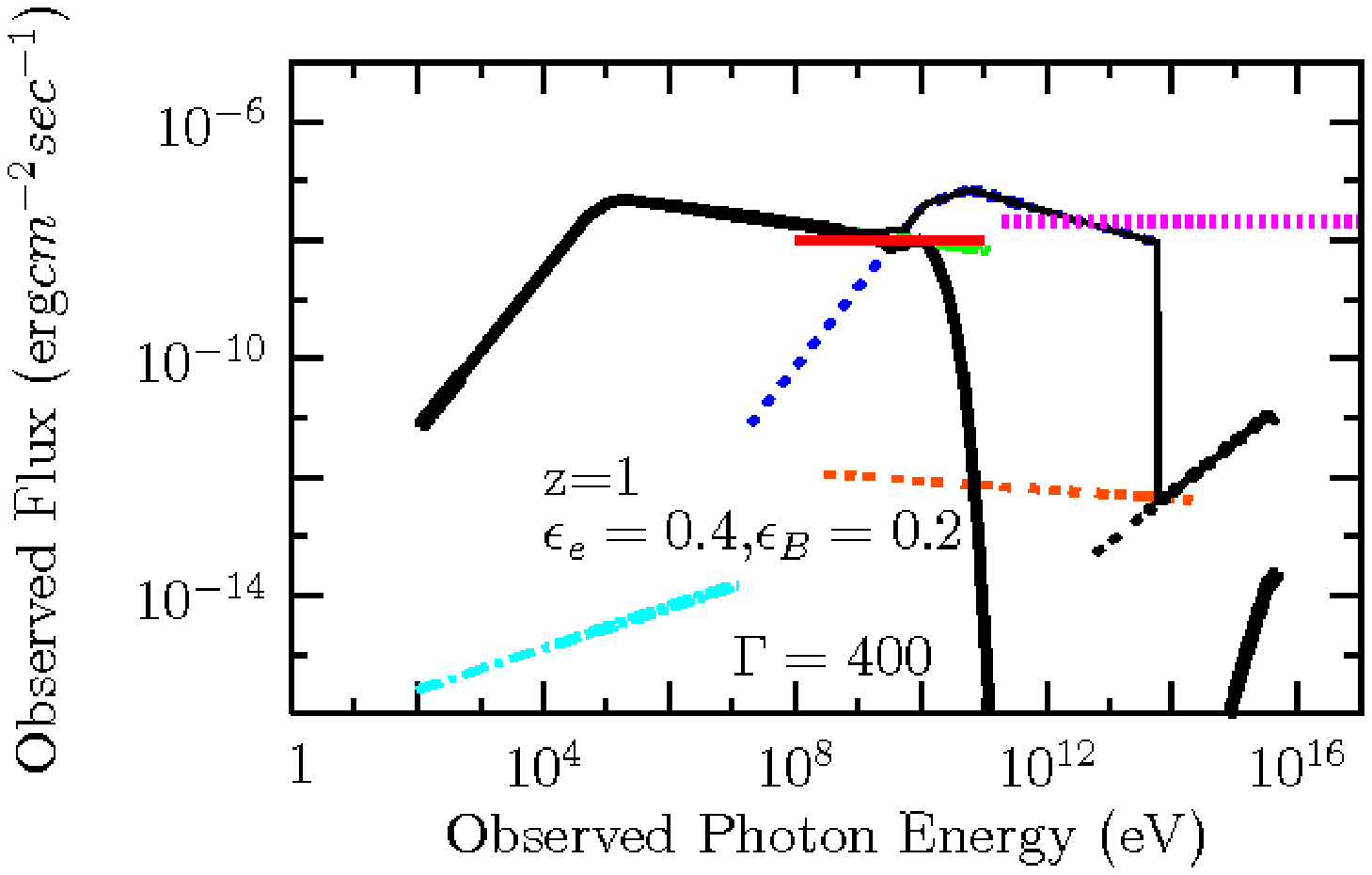}
\includegraphics[width=10cm]{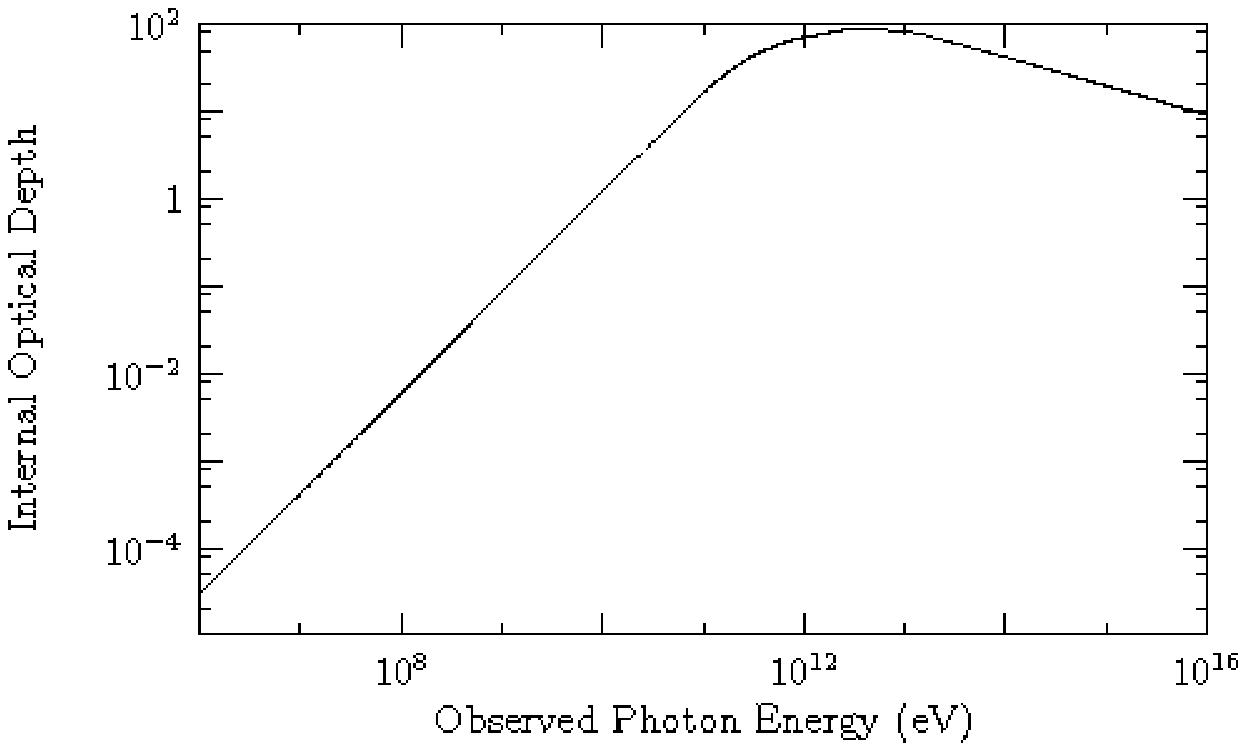}
\caption{A leptonic-component-dominated slow cooling spectrum. 
(a) The different components 
of the photon energy spectrum from the internal 
shocks for the following parameters in the slow-cooling regime:  $E_{iso}=10^{53}$erg, 
$L_{iso}=10^{52}$erg/s, $t_v=0.01$s and $f_c=2500$. The thick solid black curve 
represents the final spectrum after including the effect of internal optical 
depths. The thin solid black curve represents the synthesized spectrum before 
including the effect of internal opical depths. The long dashed (green) curve is the 
electron synchrotron component; the short dashed (blue) curve is the electron IC
component; the double short dashed (black) curve on the right side is for $\pi^0$ 
decay component; the triple short dashed (orange) line represents the synchrotron 
radiation produced by positrons generated in $\pi^+$ decays; the dash-dotted 
(light blue) line represents the proton synchrotron component. The tiny red horizontal 
line between $10^{8}$ and $10^{11}$eV represents GLAST's threshold. The pink dotted 
horizontal line above $2\times10^{11}$eV represents the sensitivity of VERITAS 
experiment (b) Internal optical depths plotted against energy 
for the parameters adopted in (a).} 
\label{slow_c_1}
\end{center}
\end{figure}

We vary the values of the equipartition parameters ($\epsilon_e$, $\epsilon_B$, 
$\epsilon_p$) and study the variations in the photon energy fluxes generated by 
various processes. The emission level of the electron IC spectral component 
decreases with decreasing $\epsilon_e$ (fixing $\epsilon_B$) since $Y_e$ is 
decreasing. Moreover, as we decrease $\epsilon_e$ the minimum injection energy 
of electrons $E_{\gamma,m}$ also decreases. In the slow cooling regimes, it is 
$E_{\gamma,c}$ that defines the peak energy in the electron synchrotron spectrum, 
which could be adjusted to the sub-MeV range by adopting a suitable $f_c$ value. 
The change of $E_{\gamma,m}$ therefore mainly affects the calculated internal 
optical depth.

By lowering $\epsilon_e$, we check the parameter regime where the hadronic 
component becomes comparable. Since eletrons are much more efficient emitters than
protons, the parameter regime for the hadronic component to be 
comparable to the leptonic component in the high energy regime is 
$\epsilon_e/\epsilon_p \sim m_e/m_p < 10^{-3}$.\footnote{Proton energy loss and 
their contribution to high energy photon emission in the early afterglow phase 
has been studied earlier by \citet{peer5}. Our results for the prompt emission phase 
are generally consistent with them. In order for the proton synchrotron component
to be significant, even smaller $\epsilon_e$
(than $10^{-3}$) is demanded. Considering that photon-pion emission is more
efficient than proton synchrotron emission, the condition $\epsilon_e/\epsilon_p 
\sim m_e/m_p < 10^{-3}$ can allow the hadronic components to be comparable to 
(but not dominant over) the leptonic components.} A similar conclusion has
been drawn for the external shocks (Zhang \& M\'esz\'aros 2001). In Fig.2,
with $\epsilon_e =10^{-3}$, $\epsilon_B=0.05$ and $\epsilon_p=0.849$. In order 
to adjust $E_{\gamma,c}$ to the sub-MeV range, $f_c=50000$ is needed. In order 
to match the observed MeV emission flux by electron synchrotron, a large energy 
budget is needed due to a small $\epsilon_e$: $E_{iso}=10^{56}$ ergs and $L_{iso}
=10^{55}~{\rm erg~s^{-1}}$. Such a large energy budget has been suggested
before \citep{tot1}, but afterglow observations and modeling in the pre-Swift
era have generally disfavored such a possibility \citep{pan3}. In the Swift
era, however, a large afterglow kinetic energy for some GRBs is not ruled out.
For example, the bright afterglow of GRB 061007 demands a huge kinetic energy
if the afterglow is produced by isotropic external shocks  
\citep{mun,schady}. Modeling some X-ray afterglows below the cooling frequency
requires a low $\epsilon_B$ and/or a large afterglow kinetic energy at least 
for some GRBs \citep{bing2}. We therefore still consider such a possibility.
In Fig.2, the break energy in the photon energy spectrum due to the minimum 
injection energy of electrons is below the synchrotron self absorption energy. 
The break energies appear in the order of $E_{\gamma,m}<E_{ssa}<E_{\gamma,c}$ 
in the synchrotron and IC electron spectra. The spectral index of the photon 
energy flux is $7/2$ between $E_{\gamma,m}$ and $E_{ssa}$, $-(p-3)/2$ between 
$E_{ssa}$ and $E_{\gamma,c}$, and $-(p-2)/2$ above $E_{\gamma,c}$. We can see
that in the TeV energy regime beyond the maximum electron synchrotron 
energy, the positron synchrotron emission from $\pi^{+}$ decay becomes dominant. 
Moreover, when $\epsilon_e$  is small, $Y_e$ is small, hence $Y_p$ becomes small.
In this case the proton synchrotron component becomes comparable to the spectral
components due to synchrotron radiation of the secondary positrons and $\pi^0$ 
decays. The internal optical depth is plotted in Fig.2b, which peaks at a higher
energy than that in Fig.1b.

\begin{figure*}
\begin{center}
\includegraphics[width=10cm,height=7cm]{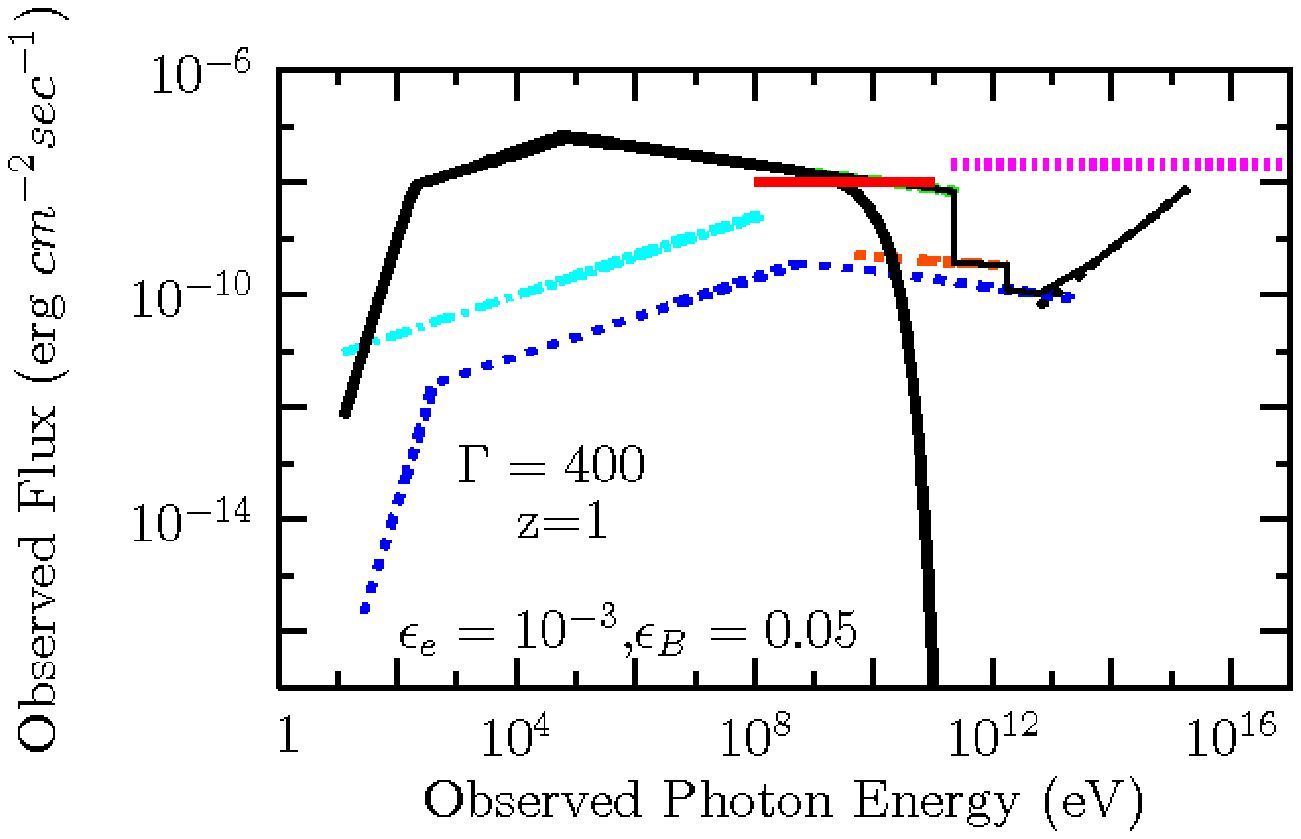}
\includegraphics[width=10cm,height=7cm]{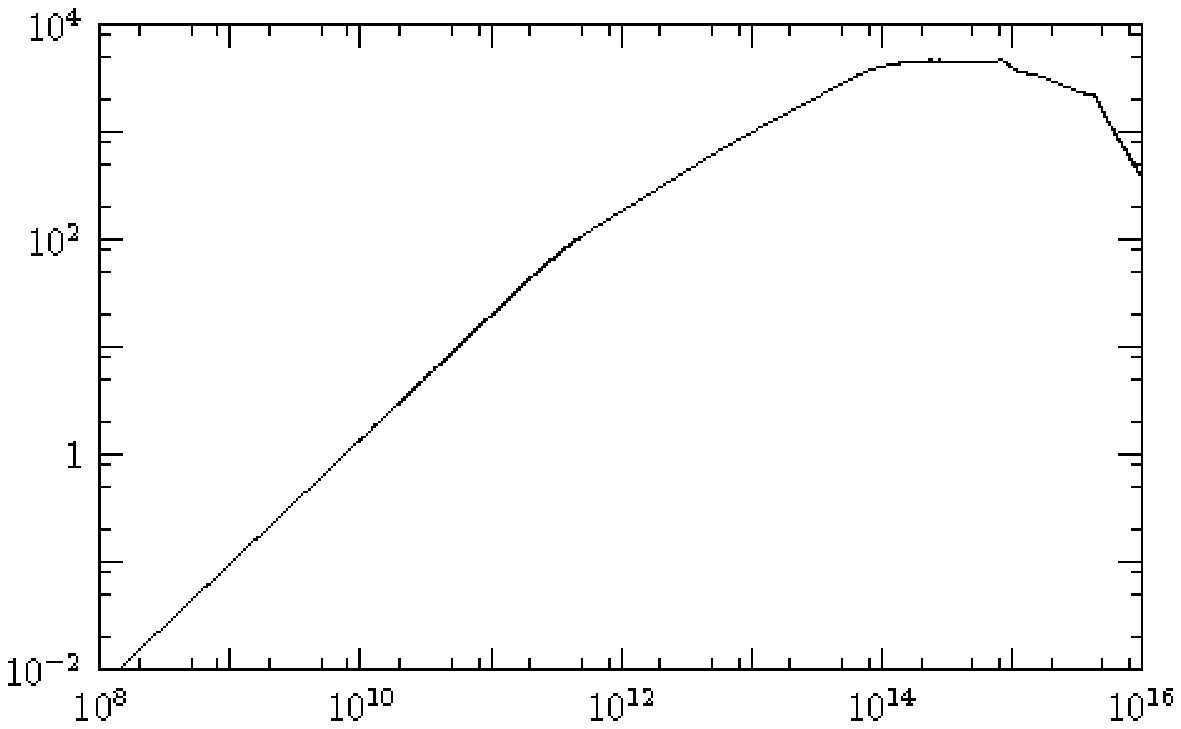}
\caption{ A slow-cooling spectrum with significant hadronic contribution
(a) The spectra of various components. Parameters: $\epsilon_e=10^{-3}$, 
$\epsilon_B=0.05$, $\epsilon_p=0.849$, $t_v=0.01$s, $f_c=50000$, $E_{iso}=10^{56}$erg 
and $L_{iso}=10^{55}$erg/s. Same line styles have been used as in Fig.1. (b) The 
corresponding internal optical depths.}
\end{center}
\end{figure*}

If the post shock magnetic field does not decay within a short distance ($f_c=1$), 
internal shocks are in the standard fast-cooling regime. We calculate such a case
in Fig.3.  The shock parameters are $\epsilon_e=0.6$, $\epsilon_B=0.2$, $L_{iso}
=10^{52}~{\rm erg~s^{-1}}$, $E_{iso}=10^{53}$ erg. In this case the break energies 
appear as in the order of $E_{C}<E_{ssa}<E_{m}$. The photon energy spectral indices 
are $13/8$, $1/2$ and $-(p-2)/2$, respectively, in the three energy regimes.  

\begin{figure*}
\begin{center}
\includegraphics[width=10cm,height=7cm]{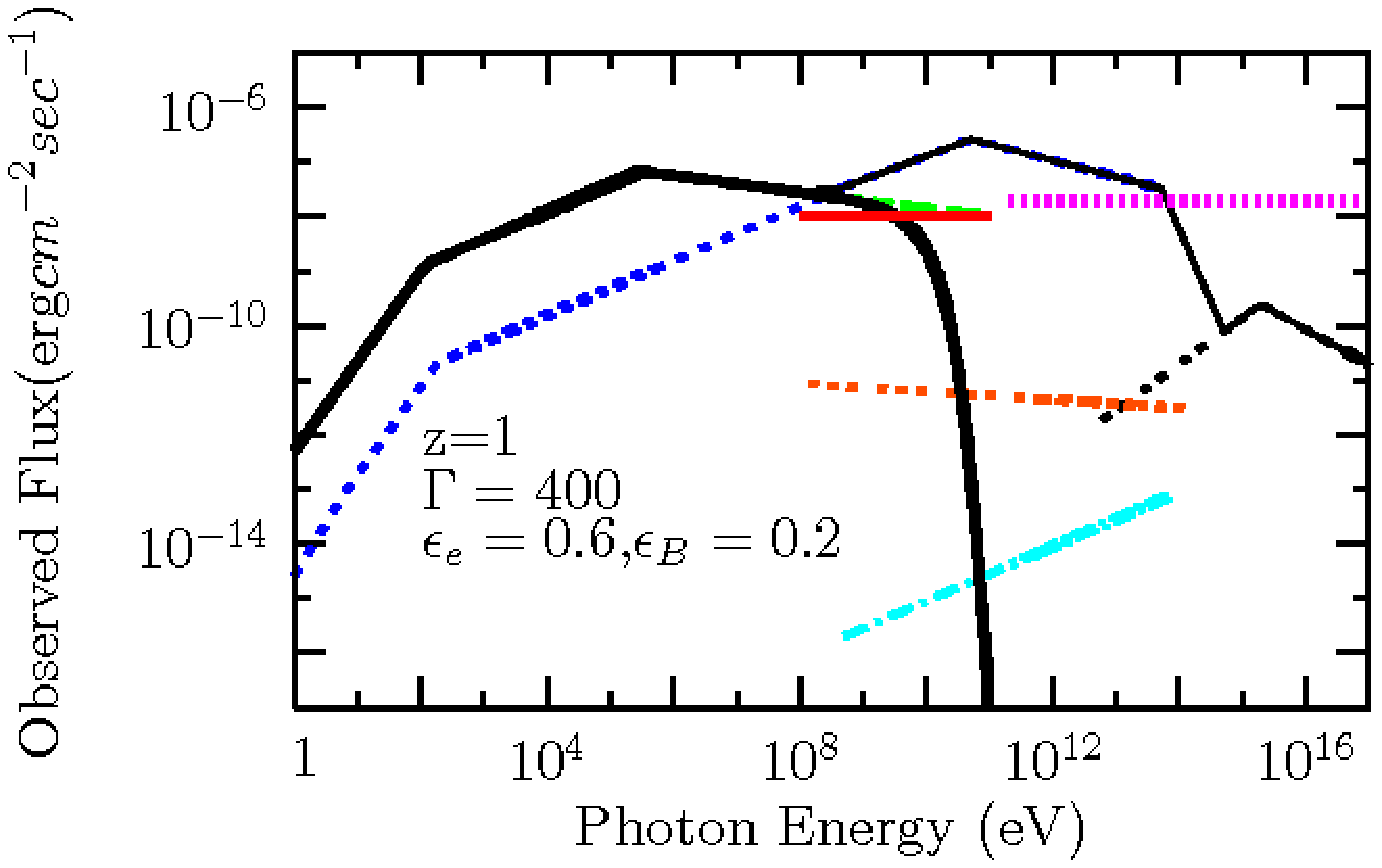}
\includegraphics[width=10cm,height=7cm]{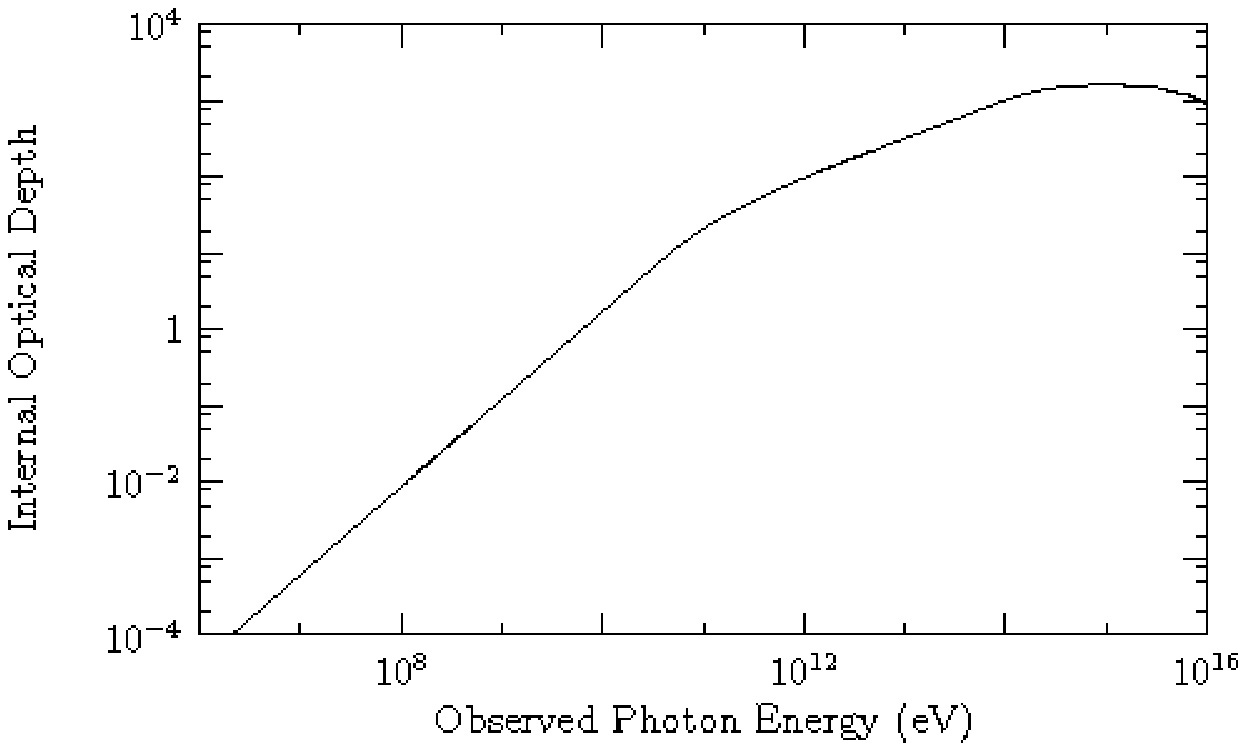}
\caption{A leptonic-component-dominated fast-cooling spectrum. (a) The spectra
of various components. Parameters: $\epsilon_e=0.6$, $\epsilon_B=0.2$,
$\epsilon_p=0.2$, $t_v=0.01$s, $f_c=1$, $E_{iso}=10^{53}$erg 
and $L_{iso}=10^{52}$erg/s. Same line styles have been used as in Fig.1. (b) The 
corresponding internal optical depths.}
\end{center}
\end{figure*}
The pair opacity depends on the bulk Lorentz factor. When $\Gamma$ is large enough,
the ultra-high energy photons would have lower internal optical depth and may escape
from the internal shocks \citep{raz1}. To test this, in Fig.4, we re-calculate with 
the parameter set for Fig.1, but increase $\Gamma$ to 1000. The slow-cooling parameter
$f_c$ is adjusted to 50 to maintain the sub-MeV energy peak. The results indeed
suggest that the attenuation of the high energy photons is weaker.

\begin{figure*}
\begin{center}
\includegraphics[width=10cm,height=7cm]{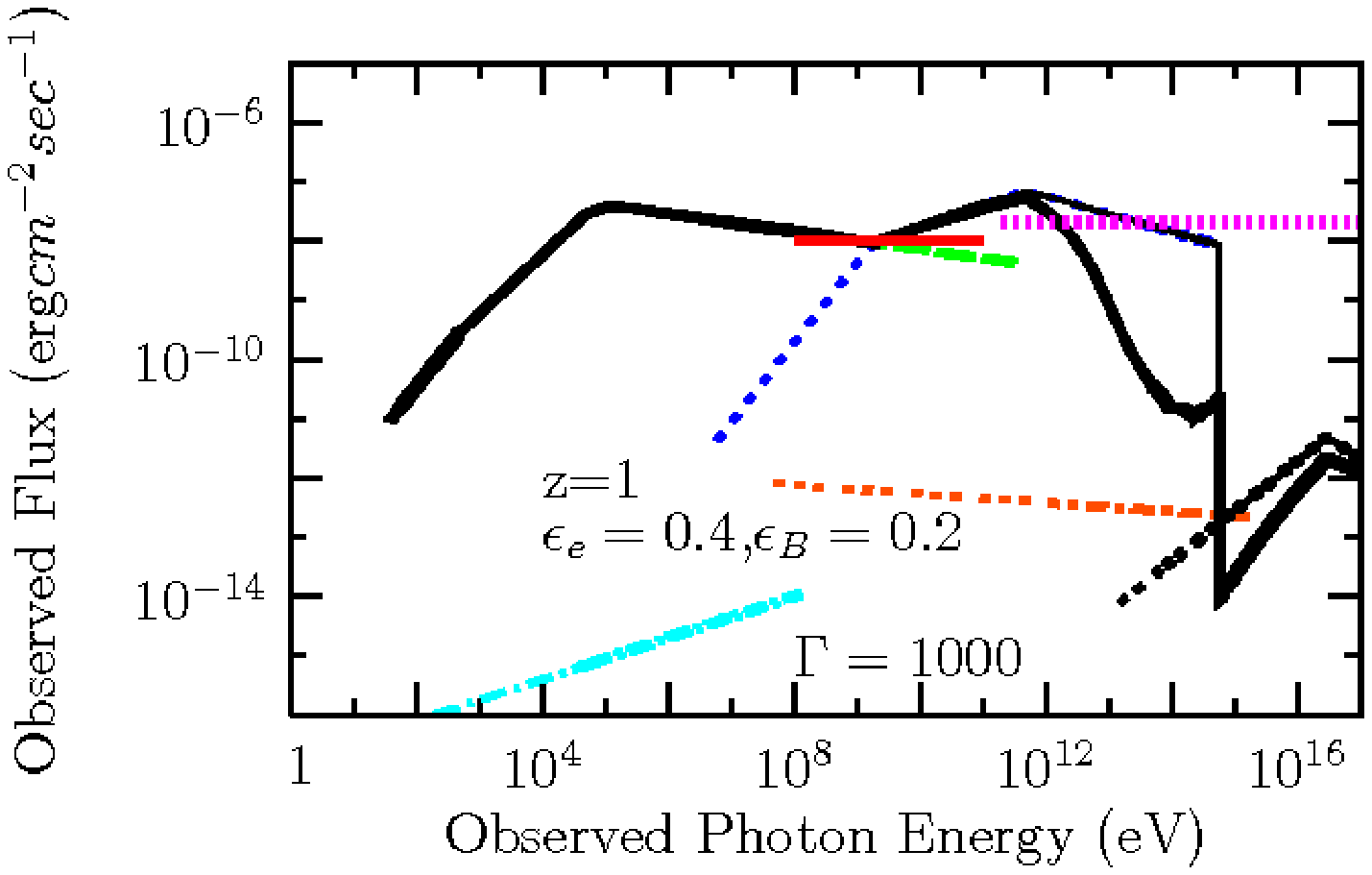}
\includegraphics[width=10cm,height=7cm]{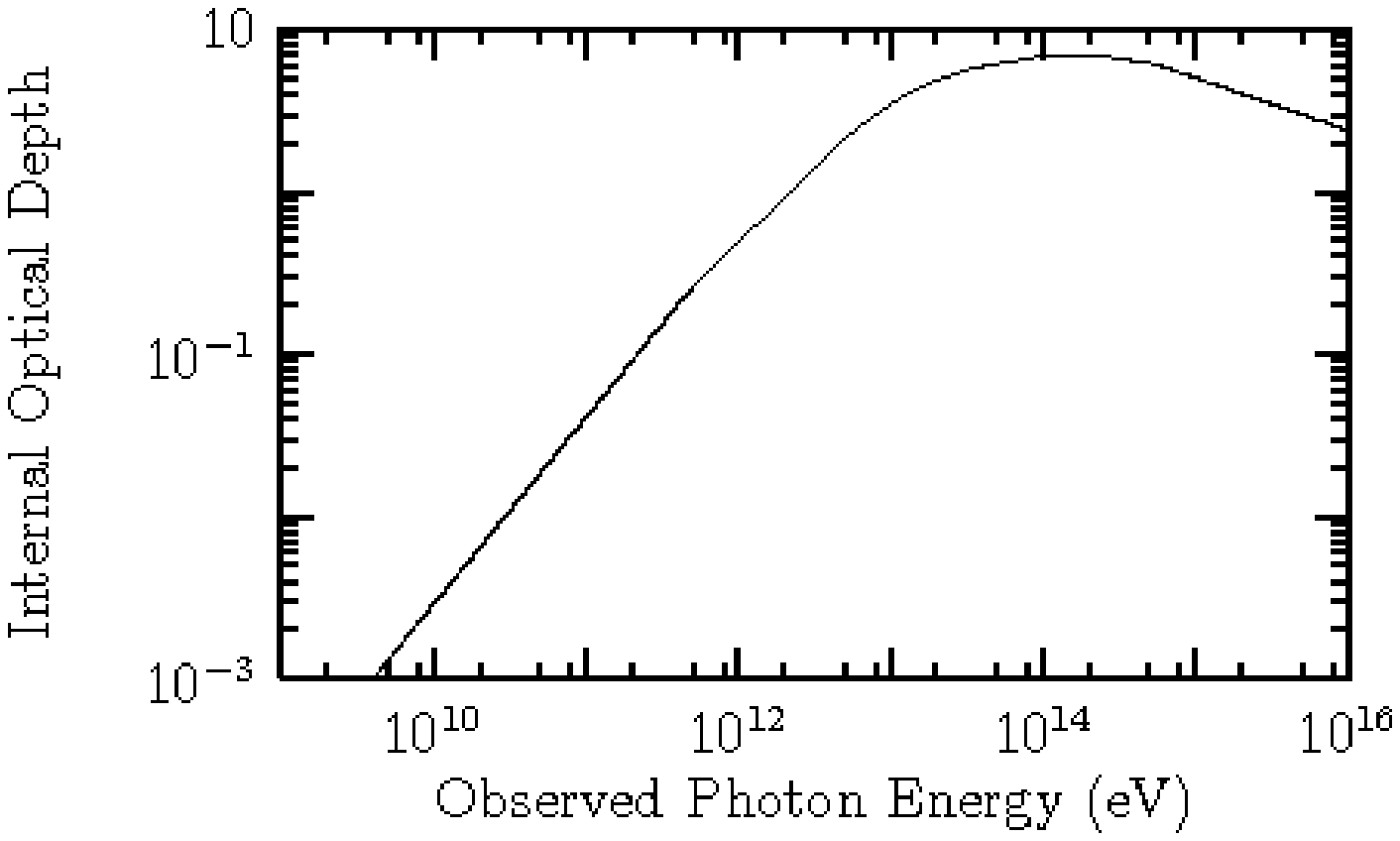}
\caption{The case of a higher Lorentz factor. (a) The spectra
of various components. Parameters: $\Gamma=1000$ and $f_c=50$. All the other parameters 
are the same as in Fig.1. (b) The corresponding internal optical depths.}
\end{center}
\end{figure*}

The observational breakthough in 2005 suggests that at least some short GRBs are
low-fluence, nearby events that have a distinct progenitor than long GRBs
(Gehrels et al. 2005; Bloom et al. 2006; Fox et al. 2005; Villasenor et al. 2005; 
Barthelmy et al. 2005; Berger et al. 2005). To check the prospect of detecting
short GRB prompt emission with high energy detectors such as GLAST, we perform
a calculation for the parameters of a short GRB in Fig.5. Due to their short
durations, short GRB detections are favorable for high luminosity and relatively
``long durations''. We therefore take an optimistic set of parameters with 
$L_{iso}=10^{51}~{\rm erg~s^{-1}}$, $T_{90}=1$ s, and $z=0.1$. Other parameters
include: $\Gamma=800$, $t_v=1$ ms, $\epsilon_e = 0.4$, $\epsilon_B=0.2$, $\epsilon_p=0.4$,
$f_c=50$. The photon flux from synchrotron radiation of electrons peaks at 0.1MeV. 
Fig.5a suggests that the high energy component of such a burst is
barely detectable by GLAST. The internal optical depth of this set of parameters
does not grow to very large values (maximum 10), so that the attenuation signature
is not significant in Fig.5a. The dip around several $10^{13}$ eV corresponds to
the optical depth peak, above which the attenuated flux starts to rise. The abrupt
drop at several $10^{14}$ eV corresponds to the disappearance of the electron IC
component at high energies.

\begin{figure*}
\begin{center}
\includegraphics[width=10cm,height=7cm]{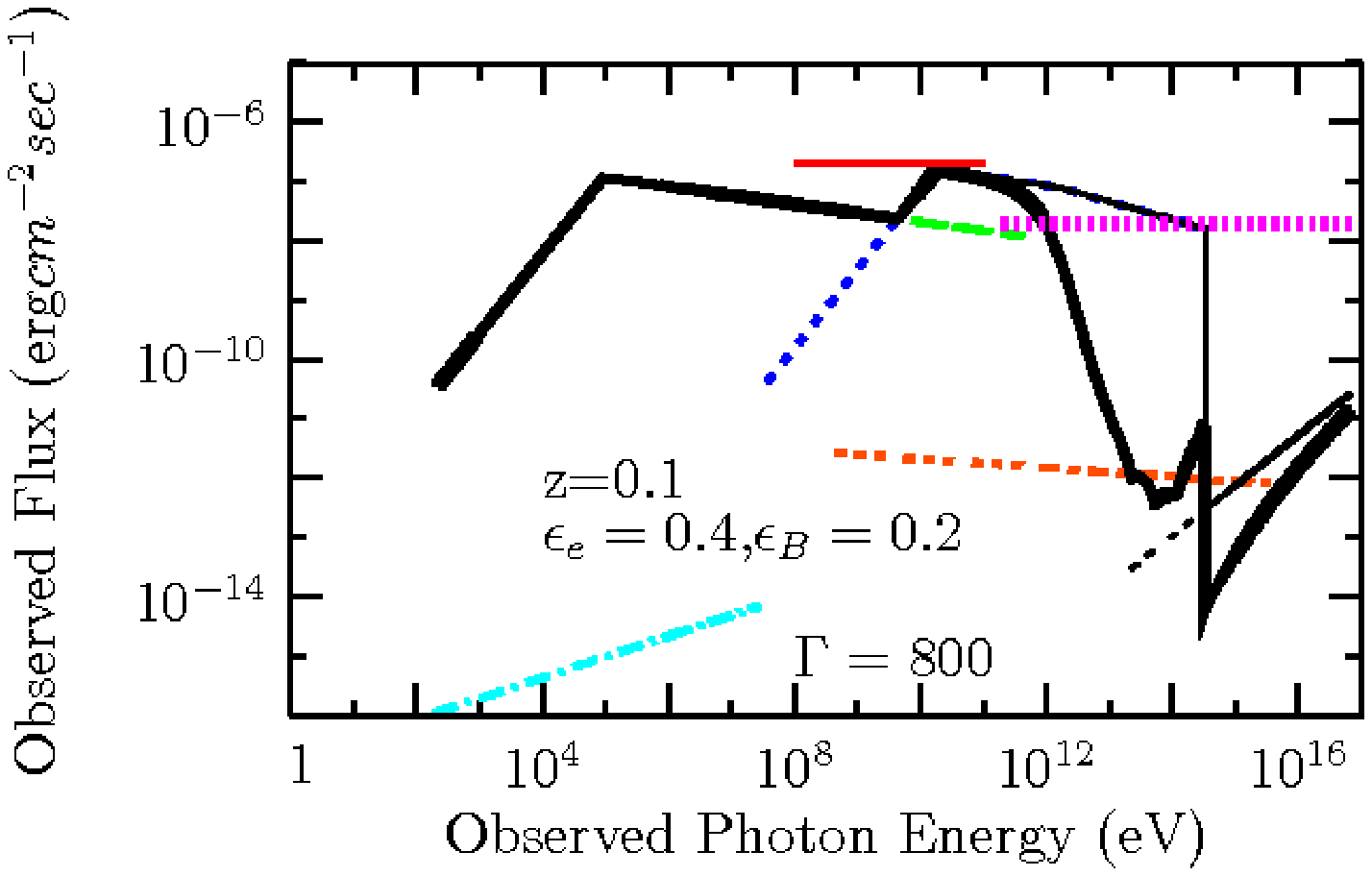}
\includegraphics[width=10cm,height=7cm]{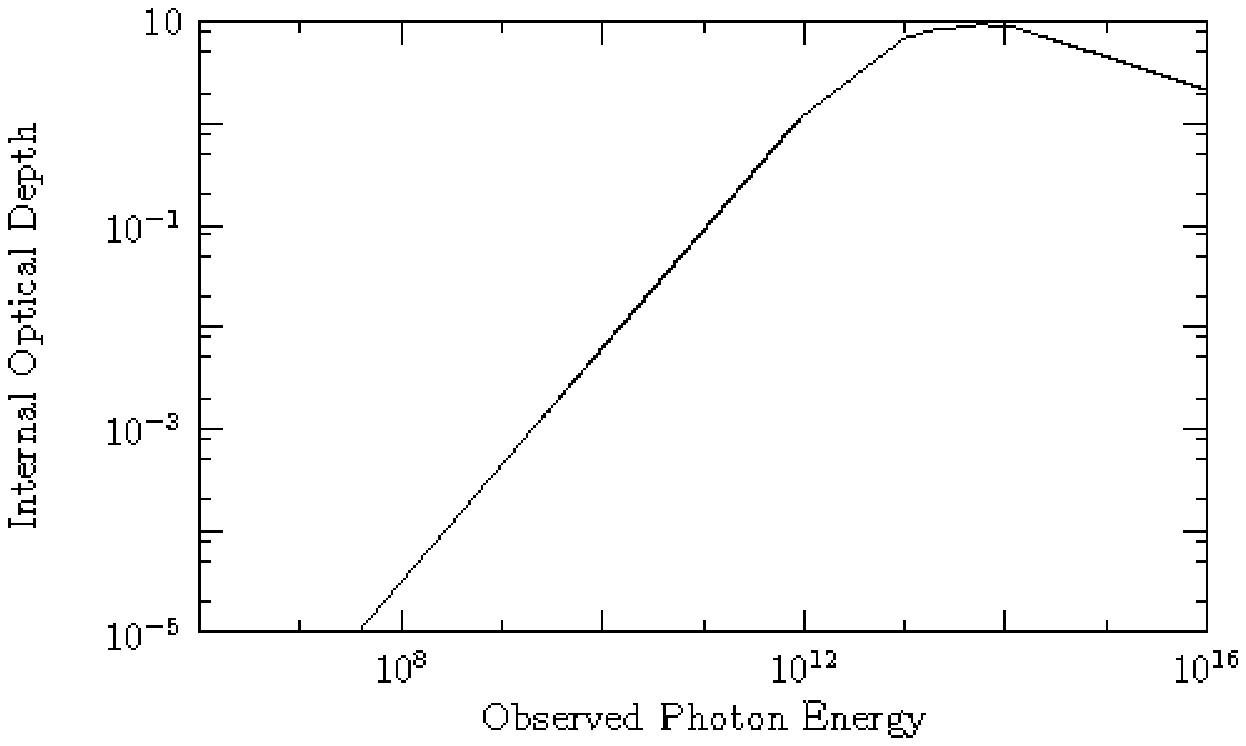}
\caption{An energetic short GRB. (a) The spectra
of various components. Parameters: $\epsilon_e=0.4$, $\epsilon_B=0.2$,
$\epsilon_p=0.4$, $t_v=0.001$s, $\Gamma=800$, $f_c=50$, $E_{iso}=10^{51}$erg 
and $L_{iso}=10^{51}$erg/s. Same line styles have been used as in Fig.1. (b) The 
corresponding internal optical depths.}
\end{center}
\end{figure*}

\section[]{Conclusions and Discussion}

We have calculated the broad-band spectrum of GRBs from internal shocks
for a wide range of parameter regimes. We did not take into account the external
attenuation of TeV photons by the infrared radiation background and that of the 
PeV photons by the cosmic microwave background. These external processes would 
further attenuate our calculated spectrum in high energy regimes, and reprocess 
the energy to delayed diffuse emission \citep{dai,stecker,wang4,raz1,casa,murase}. 
Such processes are not relevant for most of the calculations presented, however, 
since the internal attenuation already cuts the observed spectrum below TeV. They 
are however important for high Lorentz factor cases in which more high energy
photons are leaked out of the internal shock region. The external attenuation is 
also prominant for high energy emission from the external reverse/forward shocks
and the external IC processes related to X-ray flares. These processes have
been extensively discussed in other papers (referenced in Introduction) and 
they are not discussed in this paper. For nearby GRBs (e.g. $z<0.3$), TeV emission
is transparent. It is possible that ground-based Cherenkov detectors such as
VERITAS, Milagro would detect TeV gamma-rays from nearby energetic GRBs.

In previous treatments of hadronic components from internal shocks \citep{frag,bhat}, 
the shock accelerated protons are assumed to carry $m_p/m_e$ times more energy than 
electrons. This effectively fixed $\epsilon_e \sim m_e/m_p$, which is not justified
from the first principle. In this paper we have taken all the equipartition parameters
$\epsilon_e$, $\epsilon_p$ and $\epsilon_B$ as free parameters, and explore the
relative importance of various components in different parameter regimes. The
dominant hadronic component emission becomes interesting only when $\epsilon_e$
is extremely small. Given the same observed level of
sub-MeV spectrum, the total energy budget of the GRB needs to be very large.
Inspecting the calculated spectra for different parameter sets (Figs.1-4), one
finds that there is no clean picture to test the leptonic vs. hadronic
origin of the gamma-rays. Such an issue may be however addressed by
collecting both prompt and afterglow data. A moderate-to-high radiative
efficiency would suggest a leptonic origin of high energy photons, while
a GRB with an extremely low radiative efficiency but an extended high 
energy emission component would be consistent with (but not a proof for)
the hadronic origin.

The prompt emission produced by leptons including the effect of pair production 
has been discussed by \citet{peer1,peer2}. They calculated the emergent photon 
spectra for GRBs located at $z=1$. The lower cut-off energy in the photon flux 
produced by leptons is determined by the synchrotron self absorption energy, the 
minimum injection energy or the cooling energy depending on the values of the 
various GRB parameters. Our leptonic-component-dominated cases are consistent 
with their results, although we do not explore cases with very high compactness.
If the electrons cool down 
to trans-realtivistic energies then their high energy spectrum significantly
deviates from broken power law \citep{peer2,peer4}. For our choice of values of 
the GRB parameters this effect is not important.  
\citet{raz1} estimated the internal optical depth for pair production and showed 
that at PeV energies the optical depth decreases with increasing photon energies. 
We have rederived the optical depths for values of GRB parameters. The results are
generally consistent with \citep{raz1} except that the growth of optical depth 
with increasing energy is more gradual before the optical depth peak. This is 
a result of including the whole low energy photon spectrum (rather than the
threshold energy photons) for calculating the pair production optical depth.
The optical depths 
depend on the cross section of $\gamma\gamma$ interactions, the low energy photon 
spectra, the various break energies in those spectra, luminosities, variability 
times and the GRB Lorentz factors. A change in values of any of these parameters 
may affect the values of the optical depths at various energies. For high bulk 
Lorentz factors, the $\pi^0$ component may appear in the final 
spectra due to the reduced optical depths around PeV energy. However, these
ultra-high energy photons will be immediately absorbed in the GRB neighborhood by 
cosmic microwave photons \citep{stecker}. The reradiated energy by the $e^{+}e^{-}$ 
pairs would nonetheless contribute to the diffuse high energy $\gamma$-ray background
\citep{casa}. 

Upcoming $\gamma$-ray detectors have a good chance of detecting prompt emission 
from GRBs and reveal their physical nature during the prompt phase. Detection of
the hadronic components is difficult but it would be possible to infer the dominance 
of these components by a coordinated broadband observational campaign if they are
indeed important. More generally, detection or non detection of high energy photons 
in the prompt phase would constrain the values of various GRB parameters. In
particular, the pair attenuation feature would help to constrain the bulk Lorentz
factor of the fireball. Compared with EGRET, GLAST has a 10 times larger collecting
area and a larger field of view. It is expected that GLAST LAT would detect high
energy emission from a large number of bursts (mostly long GRBs and some bright,
relatively ``long'' short GRBs), which will open a new era of studying GRBs in the
GeV-TeV regime. On the other hand, it is difficult for VERITAS to detect prompt
high energy gamma-rays even under the most optimistic conditions. High energy 
emissions from the external shock at the early afterglow phase for nearby GRBs may be 
the better targets for VERITAS and other TeV detectors.
\\
  
We thank Brenda Dingus, Deirdre Horan, Enwei Liang, Peter M\'esz\'aros, Jay Norris, 
Asaf Pe'er, Soeb Razzaque, and Dave Thompson for useful discussion/comments and/or 
technical support. We also thank the anonymous referee for a detailed report with 
important suggestions and comments. This work is supported by NASA under grants 
NNG05GC22G, NNG06GH62G and NNX07AJ66G.

\end{document}